\documentclass[article,lettersize,journal]{IEEEtran}
\usepackage{amsmath,amsfonts}
\usepackage{amssymb}
\usepackage{algorithm}
\usepackage{array}
\usepackage[caption=false,font=normalsize,labelfont=sf,textfont=sf]{subfig}
\usepackage{textcomp}
\usepackage{stfloats}
\usepackage{url}
\usepackage{verbatim}
\usepackage{graphicx}
\usepackage{cite}
\usepackage{booktabs}
\usepackage{multirow}
\usepackage{multicol}
\usepackage{algpseudocode}
\usepackage{xcolor}
\usepackage[colorlinks=true,citecolor=blue]{hyperref}
\usepackage{pifont}
\usepackage{url}

\begin{document}

\title{Physics-Driven Semantic Scattering Structure Understanding of Aircraft Target in SAR Images}
\author{Yifei~Yin, Xiaogang~Yu*, Hao~Shi*, Liang~Chen,
Wei~Li~\IEEEmembership{Senior Member,~IEEE}%<-this % stops a space

% \author{Yifei~Yin, Xiaogang Yu, Hao Shi, Liang Chen and Wei Li%
\thanks{This work was supported by the Fundamental and Interdisciplinary Disciplines Breakthrough Plan of the Ministry of Education of China No. JYB2025XDXM115.}% <-this % stops a space
\thanks{Y. Yin, H. Shi, L. Chen, and W. Li are with the School of Information and Electronics, Beijing Institute of Technology, Beijing 100081, China, and with the National Key Laboratory of Science and Technology on Space-Born Intelligent Information Processing.
(Corresponding author: Hao Shi, Xiaogang Yu).}
\thanks{X. Yu is with the Beijing Institute of Remote Sensing Information, Beijing, China.}
}
 
\markboth{}%
{Yin \MakeLowercase{\textit{et al.}}: Physics-Driven Semantic Scattering Structure Understanding of Aircraft Target in SAR Images}

\maketitle
\begin{abstract}

Synthetic aperture radar (SAR) has become indispensable for target interpretation owing to its all-day and all-weather observation capability. In SAR target interpretation, electromagnetic scattering information provides a physically grounded cue beyond visual texture and has been widely exploited for target interpretation. However, existing methods remain dominated by local scattering center representations. Such unordered and component-agnostic representations are highly unstable for aircraft targets. As a result, physically existing components with weak scattering responses are often missed, resulting in the incomplete reconstructed topology structure. To address this limitation, we establish Semantic Scattering Structure Understanding as a new paradigm for SAR aircraft interpretation. Semantic scattering keypoints are defined to associate local electromagnetic responses with physically meaningful aircraft components, while visibility-aware attributes are introduced to retain weakly observable yet physically existed components. The keypoints are further organized into a stable semantic scattering structure. Build upon this, we propose S³U-SAR, a physics-driven framework to localize semantic scattering keypoints and construct the complete representation constrained by multi-dimensional physical priors containing scattering heterogeneity, rigid-body topology, speckle uncertainty. A confidence-gated joint supervision strategy is further introduced to alleviate optimization conflicts. We construct KP-SAR-Aircraft-1.0, the first fine-grained benchmark for semantic scattering structure understanding. Extensive experiments demonstrate that S³U-SAR achieves the best performance compared with baselines. Cross-category and cross-dataset evaluations further verify its robustness and transferability. Notably, in downstream orientation estimation, the method improves $\text{P}_{1^\circ}$ and $\text{P}_{5^\circ}$ by 17.07\% and 19.31\% over the strongest baseline. These results demonstrate we establish a new aircraft target representation paradigm that advances SAR target understanding from implicit local response regression toward explicit global structural representation. The trained model and dataset will be available at https://github.com/YYF121/S3U-SAR.
\end{abstract}


\section{Introduction}
\label{introduction}
%% Labels are used to cross-reference an item using \ref command.
Synthetic Aperture Radar (SAR) actively transmitting electromagnetic pulses and receiving the backscattered echoes to perform imaging process, capable of all-day, and all-weather day observation\cite{Zhu2021DeepLearningSAR,Chen2026SelfSupervisedDespeckling,He2025DOGAN}. Consequently, SAR images has been widely used in target detection, recognition, orientation estimation, and scene interpretation \cite{Yin2025ShipDetectionTransformer, Liu2026ATRNetSTAR, Wang2024RLOrientation}. 

Driven by advancements in deep learning, visual models based on Convolutional Neural Networks (CNNs) and Transformers have achieved tremendous success in target-level SAR image interpretation by learning discriminative visual representations from large-scale data \cite{Chen2025PGMNet, Wang2024CVSARDet}. Nevertheless, despite their empirical success, purely data-driven models still face intrinsic limitations when interpreting targets with complex physical structures. Distinct from the optical images, where target appearance is primarily characterized by continuous texture and contour cues, targets in the SAR images manifest as a collection of discrete scattering centers due to the unique coherent imaging mechanism \cite{Potter1995GTD, Potter1997ASC}. Consequently, relying exclusively on visual modality features is inadequate to accurately capture the intrinsic physical attributes.

To address this, foundational research established the Attributed Scattering Center (ASC) framework for Automatic Target Recognition (ATR), systematically demonstrating that man-made targets in SAR images are dominated by isolated physical scatterers rather than continuous optical textures \cite{Potter1997ASC}. Building upon this physical perspective, pioneering advancements have effectively addressed the challenges of severe occlusion and structural articulation by leveraging the invariant topological configurations of these discrete scattering centers\cite{Jones1999ArticulatedOccluded}. In recent years, explicitly integrating electromagnetic scattering characteristics into deep neural networks has emerged as a prevailing trend to overcome the limitations of purely data-driven visual models. FEC \cite{Zhang2021FEC} focused on feature-level integration by parameterizing extracted ASCs into a bag of visual words for concatenation with CNN representations. To enhance physical interpretability, PAN \cite{Feng2023PAN} and ASC-SepNet \cite{Zhang2025ASCSepNet} utilize scaled dot-product attention and dual-driven separability, respectively, to actively reconstruct and align local electromagnetic priors with global visual feature maps. SARATR-X \cite{Li2025SARATRX} establishes a customized SAR foundation model that captures highly generalizable physical-visual representations. Scattering Prompt Tuning (SPT) \cite{Guo2024SPT} transforms discrete scattering attributes into text prompts, enabling the cross-modal fusion of electromagnetic mechanisms and visual features within the input space. These studies demonstrate that electromagnetic knowledge is crucial for improving SAR target representation.

\begin{figure}[h]
\centering
\includegraphics[width=\columnwidth]{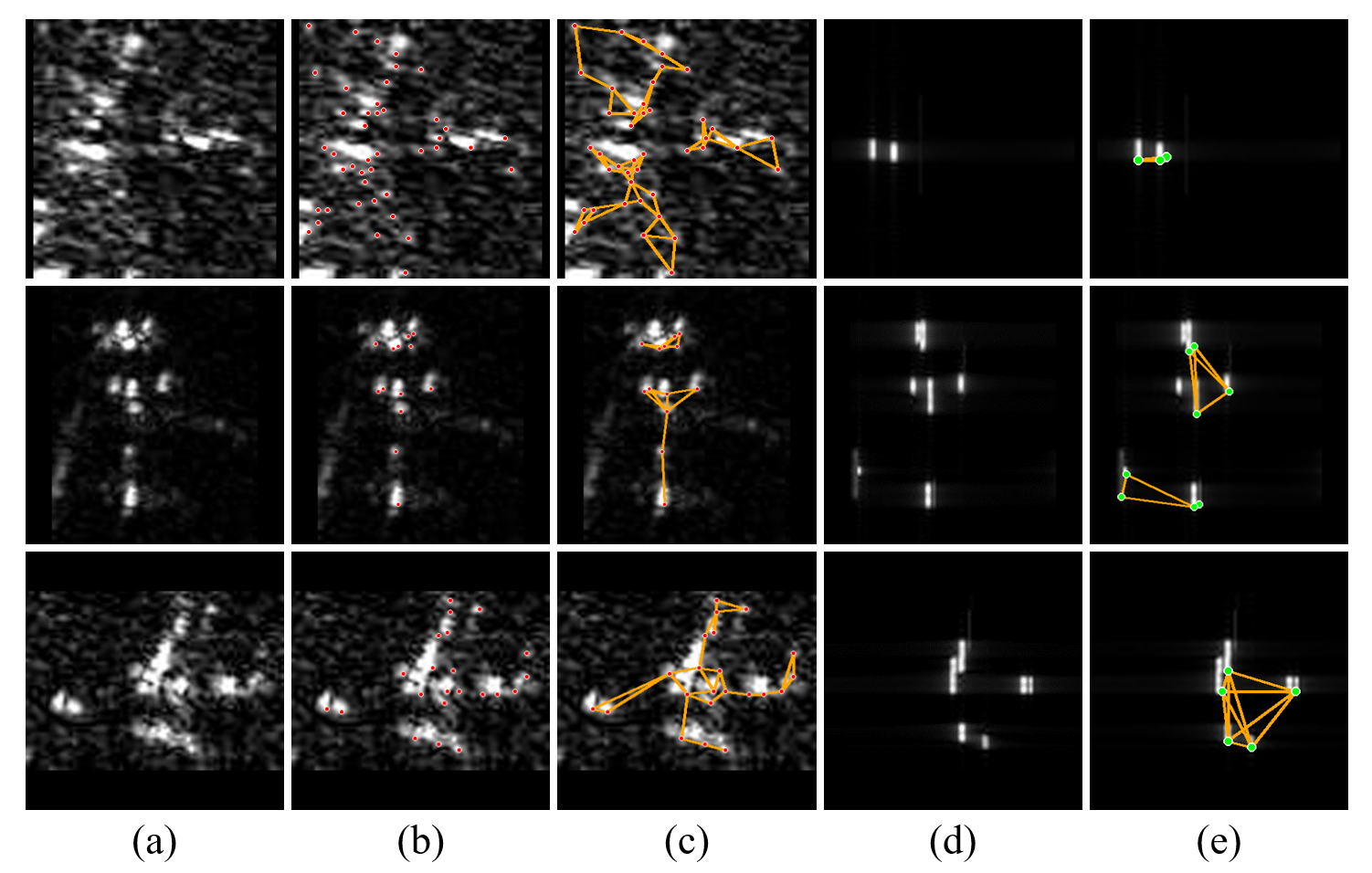}
\caption{Examples of SC- and ASC-based representations of SAR aircraft targets. From left to right: (a) original image, (b) extracted SCs, (c) topology constructed from SCs, (d) extracted ASCs, and (e) topology constructed from ASCs.}
\label{fig:intro_sc_asc}
\end{figure}

However, existing methods remain fundamentally limited for fine-grained aircraft understanding. Conventional Scattering Centers (SCs) or Attributed Scattering Centers (ASCs) are essentially unordered and component-agnostic\cite{Gerry1999ParametricSAR, Liao2024EMINet}. The semantic correspondence between scattering response with physical component could not be explicitly encoded \cite{Li2022GCN}. In addition, distinct from vessels and vehicles with relatively regular morphological profiles, aircraft targets are characterized by highly intricate spatial geometric configurations\cite{Kang2023STNet}. Under complex imaging conditions, structural components located away from the radar line-of-sight frequently exhibit weak or entirely absent scattering responses, accompanied by drastic morphological variations across different viewing angles. As illustrated in Fig.~\ref{fig:intro_sc_asc}, conventional extraction methods are highly susceptible to coherent speckle noise and consistently fail to capture these physically existent but visually weak scattering centers. The absence of these essential nodes results in severe deviations of constructed topological structure from the actual rigid-body physics. Consequently, the representation of aircraft targets is displayed as unordered scattering responses, leading to incomplete, unstable, and physically inconsistent interpretations.

The above limitations reveal a critical gap in current SAR target interpretation: there is still no semantic structural representation that explicitly associates SAR scattering responses with aircraft physical components and their rigid-body topology. In other words, existing methods mainly answer the question of where scattering responses appear, but they do not sufficiently address what physical components these responses correspond to and how to organize the structure of these components. 

To address the challenge, this paper investigates a new problem termed Semantic Scattering Structure Understanding for SAR aircraft targets. The objective is to parse a SAR aircraft image into a set of physically meaningful semantic scattering keypoints and a physics-constrained structural topology. A semantic scattering keypoint denotes a key aircraft component with explicit physical semantics and is associated with characteristic electromagnetic scattering response, such as the nose, tail, wing tips, and engines. Different from conventional scattering centers, semantic scattering keypoints establish explicit correspondences between SAR scattering responses and aircraft components. Furthermore, a physics-constrained structural topology is introduced to model the rigid-body geometric relationships among these semantic keypoints. In this way, SAR aircraft representation is transformed from an unordered set of local scattering responses into a structured semantic graph grounded in both electromagnetic scattering mechanisms and aircraft physical geometry.

The proposed problem is conceptually inspired by human pose estimation (HPE) \cite{Lin2014COCO,Cao2021OpenPose, Xiao2018SimpleBaseline, Andriluka2014MPII}, where objects are represented by semantic joints and skeletal connections. In the generic vision domain, High-Resolution Network (HRNet) \cite{Sun2019HRNet} designed a approach of maintaining high-resolution representations throughout the feature extraction process, providing a strong baseline for precise pixel-level localization. Subsequently, Transformer-based models such as TransPose \cite{Yang2021TransPose} and ViTPose++ \cite{Xu2024ViTPosePP} excel in capturing long-range global contextual dependencies for joint association, while Spatial-Aware Regression \cite{Wang2024SpatialAware} integrates spatial constraints directly into coordinate regression. Additionally, generative processes DiffusionPose \cite{Wang2026DiffusionPose} and highly efficient topology networks NanoHTNet \cite{Cai2025NanoHTNet} have been designed to resolve complex joint dependencies under severe occlusions. Overall, these advancements provides a feasible framework for robustly localizing target keypoints and reconstructing topological structure.

However, directly transferring generic pose estimation methods from natural images to SAR aircraft interpretation is highly nontrivial. First, a fundamental discrepancy exists between the flexible articulated structure of the human skeleton and the rigid-body topology of aircraft. Second, the observability of SAR aircraft components varies significantly with imaging geometry. Some physically existing components may be weakly visible or even electromagnetically invisible due to side-looking observation. Meanwhile, background clutter and speckle noise may generate strong but semantically invalid responses. Third, although recent studies have advanced SAR aircraft interpretation \cite{Wang2023SARAIRcraft, Sun2022SCAN}, the absence of a standardized public benchmark with fine-grained physical annotations prevents the developments in this field.

To address the above challenges, we propose S³U-SAR, a physics-guided framework for Semantic Scattering Structure Understanding of SAR aircraft target. This work represents the first attempt to extract semantic scattering centers and reconstruct the stable topological structure of SAR aircraft targets. Rather than treating SAR aircraft as unordered scattering responses, S³U-SAR represents each target as a set of semantic scattering keypoints associated with SAR-specific visibility attributes. Based on this representation, the proposed framework integrates scattering-intensity-aware localization, rigid-body topological regularization, entropy-based speckle suppression, and confidence-gated joint supervision to achieve robust keypoint localization and structurally consistent target understanding under complex SAR imaging conditions. Extensive experiments on the proposed KP-SAR-Aircraft-1.0 benchmark, together with open-set cross-category evaluation, cross-dataset transfer, and downstream orientation estimation, demonstrate the effectiveness and generalization capability of the proposed method.

The main contributions of this paper are summarized as follows:

(1) We first introduce Semantic Scattering Structure Understanding for SAR aircraft interpretation, which aims to parse SAR aircraft targets from isolated electromagnetic responses into semantic scattering keypoints and physics-constrained structural topology.

(2) We construct KP-SAR-Aircraft-1.0, the first fine-grained benchmark for semantic scattering structure understanding in SAR images. The benchmark provides semantic keypoint annotations, SAR-specific visibility attributes, structural topology definitions, category labels, and orientation information based on Gaofen-3 aircraft images, offering a standardized foundation for future research.

(3) We establish a physics-guided semantic scattering representation for aircraft targets. Semantic scattering keypoints are defined according to aircraft physical components and electromagnetic scattering characteristics. A visibility-aware attribute is further introduced to decouple physical existence from scattering observability, allowing weak but physically existing components to be explicitly distinguished. Ultimately, a comprehensive and stable topological semantic structure is constructed, strictly governed by the rigid-body physics of the aircraft.

(4) We propose S³U-SAR, a robust framework for semantic scattering structure understanding. During the regression process, physical priors including scattering intensity, rigid-body topology, and speckle noise are introduced for constraints. Furthermore, a novel confidence-gated joint supervision strategy is designed to dynamically resolve alignment imbalance and gradient conflicts during the training phase.    

(5) We conduct comprehensive experiments to validate the effectiveness and transferability of the proposed representation approach. Extensive experiments demonstrate that S³U-SAR outperforms representative  baselines and maintains strong robustness under unseen-category and cross-dataset scenarios. Especially, a novel paradigm for orientation estimation is established, yielding an improvement of 17.07\% and 19.31\% in $\text{P}_{1^\circ}$ and $\text{P}_{5^\circ}$ over the second-best approach. 

The remainder of this paper organized as follows. Section II construct the key semantic scattering centers model. Section III discuss the proposed S$^3$U-SAR. Section IV conduct extensive experiments to demonstrate the superiority and robustness of the proposed method. Section V conclude the paper and discuss the future work.

\section{Semantic Scattering Structure Characterization}
\label{related work}

The objective of Semantic Scattering Structure Understanding is to represent a SAR aircraft target as a structured physical-semantic entity rather than an unordered set of isolated scattering responses. To this end, we formulate a semantic scattering structural representation tailored to SAR aircraft targets:
\begin{equation}
    \mathcal{S}=(\mathcal{K},\mathcal{A},\mathcal{G}),
\end{equation}

where $\mathcal{K}$ denotes semantic scattering keypoints, $\mathcal{A}$ denotes SAR-specific visibility attributes, and $\mathcal{G}$ encodes the physics-constrained structural topology among aircraft components. Compared with conventional scattering centers \cite{Potter1997ASC,Gerry1999ParametricSAR,Kang2023STNet}, the proposed formulation explicitly associates electromagnetic scattering responses with physical component semantics and rigid-body geometry.

\begin{figure}[!t]
\centering
\includegraphics[width=0.98\columnwidth]{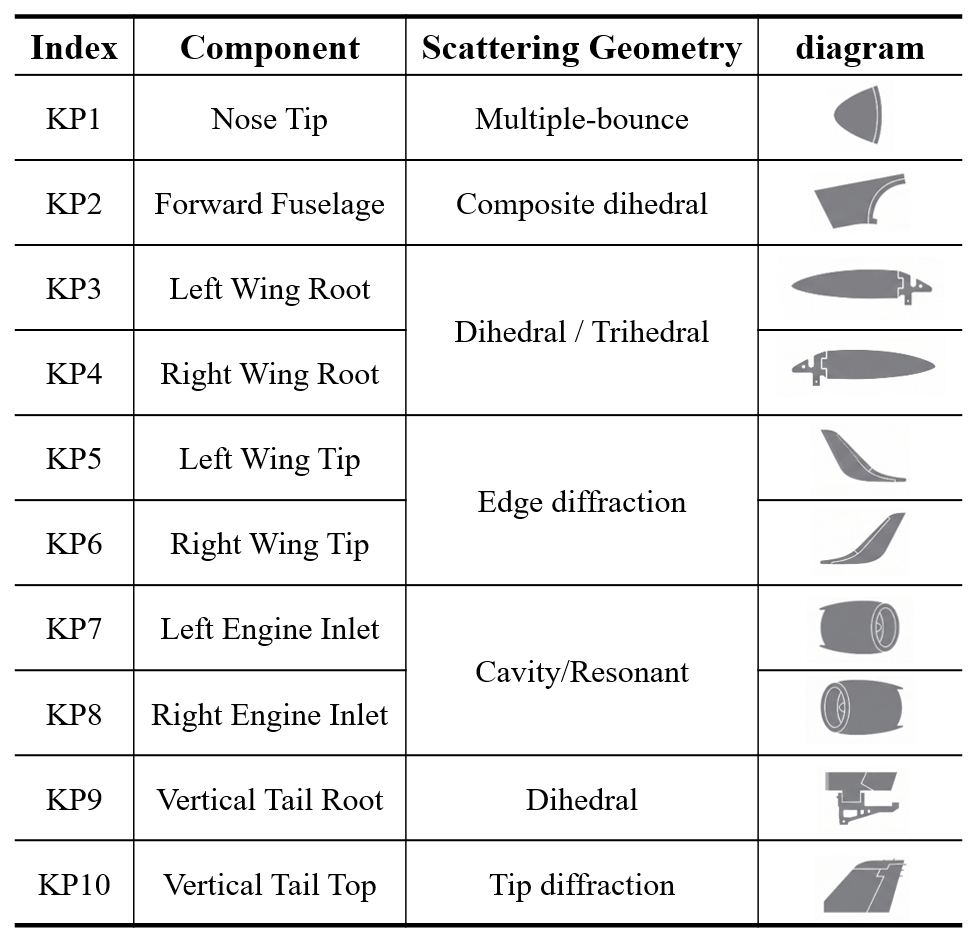}
\caption{Definition of the selected semantic scattering keypoints, including key aircraft components, corresponding scattering geometries, and schematic diagrams.}
\label{fig:keypoint_definition}
\end{figure}

\subsection{Semantic Scattering Keypoint Definition}

In SAR images, aircraft targets usually appear as sparse and discontinuous scattering responses. However, not all bright responses correspond to meaningful aircraft components, and not all physically important components necessarily produce strong scattering. Therefore, directly using local intensity peaks or extracted scattering centers as target primitives is insufficient for fine-grained aircraft understanding.

To bridge scattering responses and physical semantics, inspired by \cite{Lin2014COCO,Sun2019HRNet},  we define semantic scattering keypoints as two-dimensional projections of key aircraft components associated with electromagnetic scattering characteristics. Specifically, the selected keypoints are defined on physically meaningful aircraft components. These components exhibit stable or interpretable SAR scattering characteristics and are essential for describing the global aircraft configuration. Following the principle, ten semantic scattering keypoints are defined:

\begin{equation}
    \mathcal{K}=\{KP_i\}_{i=1}^{N}, \quad N=10,
\end{equation}

where $KP_i=(x_i,y_i,c_i)$ consists of image-plane coordinates $(x_i,y_i)$ and the semantic component label $c_i$. These keypoints cover the core structural components of aircraft \cite{Xiao2022PFFADN, Guo2020AircraftSARReview}, including the longitudinal fuselage, transverse wing structures, tail-related components, and engine-related regions. The detailed correspondence between keypoint indices, physical components, and scattering characteristics is provided in Fig.~\ref{fig:keypoint_definition}. This definition transforms SAR aircraft modeling from purely response-driven scattering observations into component-aware physical-semantic primitives.

\subsection{Visibility-Aware Scattering Semantics}
\label{sec: visibility}

\begin{figure}[!t]
\centering
\includegraphics[width=0.98\columnwidth]{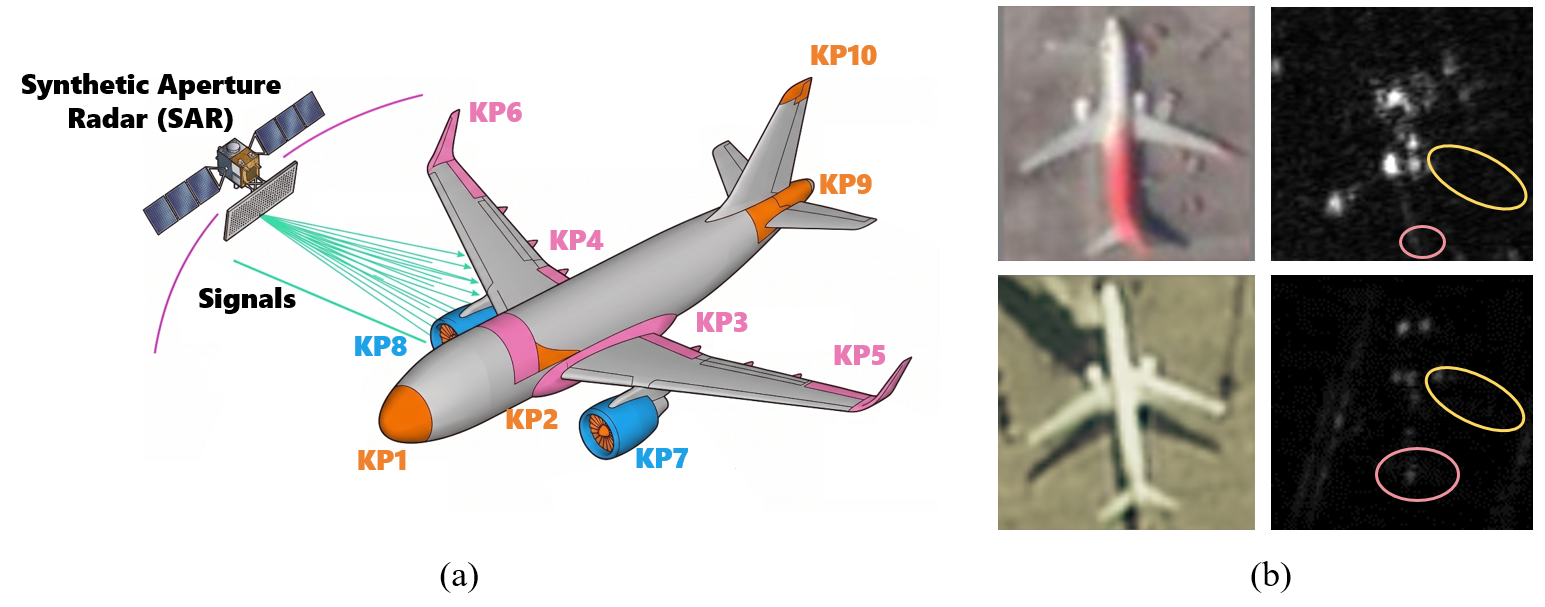}
\caption{Illustration of visibility-aware scattering semantics. (a) SAR imaging geometry and semantic scattering keypoints on aircraft components. (b) Optical and SAR image comparisons of aircraft components, where yellow ellipses indicate nearly missing scattering responses and pink ellipses indicate weak scattering responses.}
\label{fig:visibility_examples}
\end{figure}

\begin{figure}[!t]
\centering
\includegraphics[width=0.98\columnwidth]{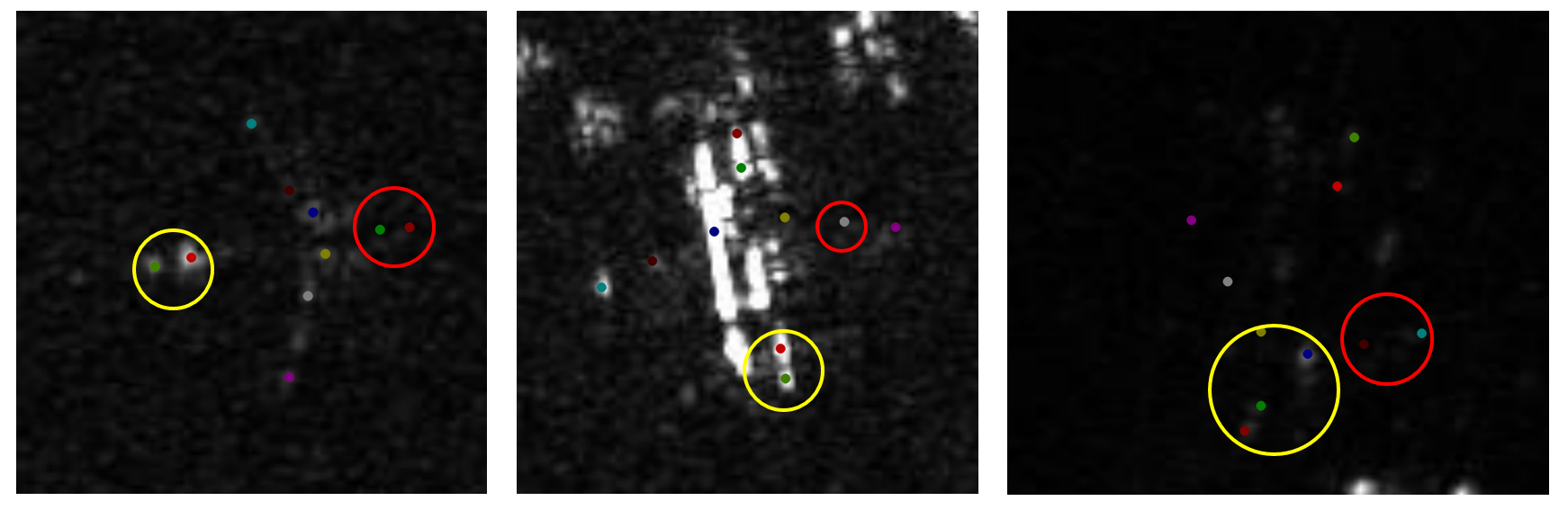}
\caption{Examples of visibility-aware semantic scattering keypoints in SAR aircraft images. Yellow circles denote scattering-salient keypoints with $v_i=2$, and red circles denote scattering-degraded keypoints with $v_i=1$.}
\label{fig:visibility_attribute}
\end{figure}

A distinctive property of SAR images is that the observability of a physical component is governed by radar imaging geometry, local target structure, material properties, and coherent imaging effects \cite{Moreira2013TutorialSAR,Oliver1998UnderstandingSAR}. As a result, a physically existing component may appear as a weak or even an invisible response, as illustrated in Fig.~\ref{fig:visibility_examples}. Forcing the network to regress components with missing or unreliable scattering response inevitably introduces substantial localization noise. The predicted structure are highly susceptible to speckle noise and background clutter \cite{Argenti2013Speckle,Zhu2021DeepLearningSAR}.

To model this ambiguity, we decouple physical existence from electromagnetic observability. For each semantic scattering keypoint, two binary attributes are introduced: semantic existence $v_i^{sem}$ and scattering observability $v_i^{scat}$. The former indicates whether the corresponding physical component belongs to the aircraft structure, while the latter indicates whether it produces a reliable scattering response in the SAR image. As illustrated in Fig.~\ref{fig:visibility_attribute}, based on these two attributes, the visibility label of $\mathcal{A}$ is defined as:
\begin{equation}
    v_i =
    \begin{cases}
    2, & v_i^{sem}=1,\; v_i^{scat}=1, \\
    1, & v_i^{sem}=1,\; v_i^{scat}=0, \\
    0, & v_i^{sem}=0.
    \end{cases}
\end{equation}

Here, $v_i=2$ denotes a scattering-salient semantic keypoint, where the corresponding physical component exists and produces reliable local scattering evidence. $v_i=1$ denotes a scattering-degraded semantic keypoint, where the component physically exists but its scattering evidence is weak, ambiguous, or unreliable due to side-looking observation, self-occlusion, or unfavorable aspect angles. $v_i=0$ denotes a semantically invalid response or irrelevant region, including bright artifacts caused by speckle noise, layover, multipath scattering, or background clutter. The proposed visibility attribute formulation prevents the representation from being dominated by local bright pixels. Weak but physically valid components are therefore retained, whereas semantically invalid bright responses are abandoned.

\subsection{Physics-Constrained Structural Topology}
\label{sec:representation}

\begin{figure}[!t]
\centering
\includegraphics[width=0.98\columnwidth]{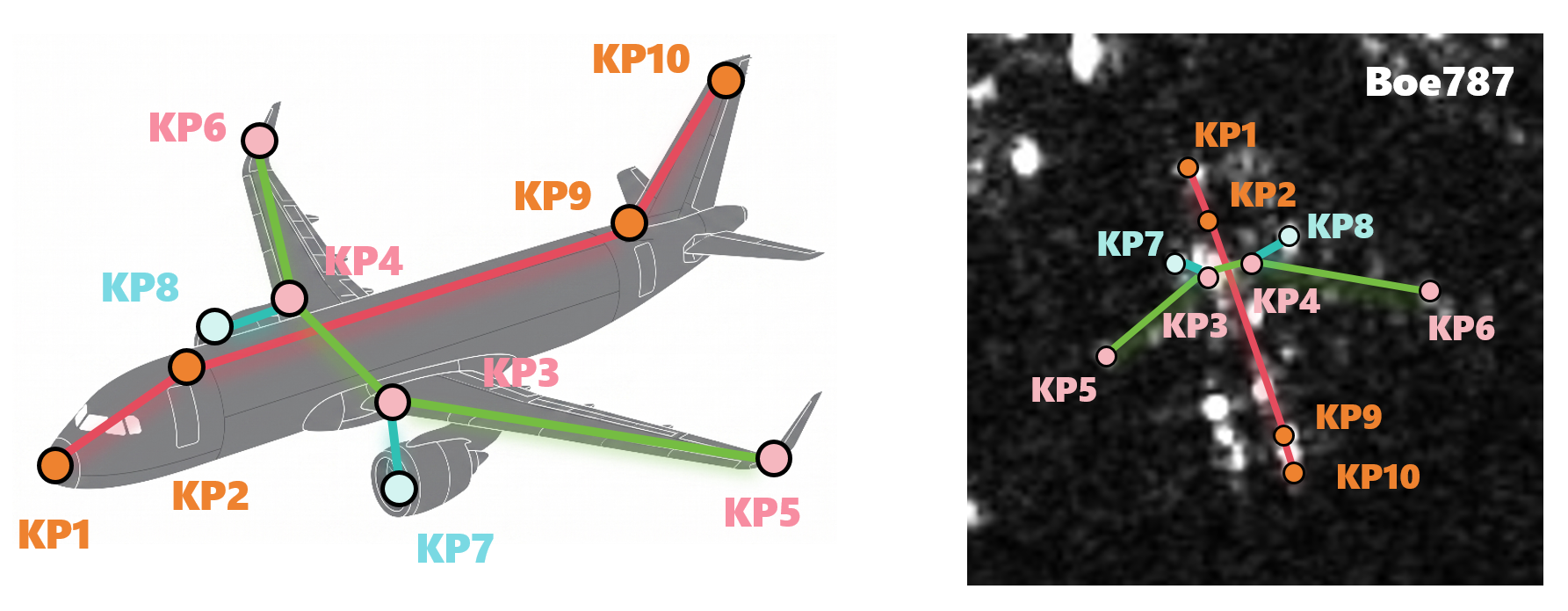}
\caption{Illustration of the physics-constrained structural topology defined on the physical aircraft structure and its corresponding mapping in a SAR aircraft image.}
\label{fig:structural_topology}
\end{figure}

Although semantic scattering keypoints provide component-level primitives, independent keypoints are insufficient to describe the global structure of aircraft targets. Aircraft are rigid-body objects with stable physical relationships among the nose, fuselage, wings, tail, and engines. These relationships remain structurally consistent across different SAR viewing geometries, even though the projected appearance and scattering intensity may vary significantly. Therefore, we construct a physics-constrained structural topology over the defined semantic scattering keypoints:
\begin{equation}
\label{equation4}
    \mathcal{G}=(\mathcal{V},\mathcal{E}),
\end{equation}

where the node set $\mathcal{V}$ corresponds to the semantic scattering keypoints in $\mathcal{K}$, and the edge set $\mathcal{E}$ encodes physically meaningful connections among aircraft components. As shown in Fig.~\ref{fig:structural_topology}, the topology is defined according to the rigid-body structure of aircraft, including the longitudinal fuselage structure, transverse wing structure, tail-related connections, and engine-related associations.

Instead of a fully connected graph, the proposed topology preserves only local and physically interpretable structural relations. The redundant long-range constraints could be avoided, which are unstable under SAR projection and aspect-angle variations. Meanwhile, the essential geometry required for aircraft structural understanding could be well retained. Moreover, the visibility-aware attributes enable structural reasoning under weak or missing scattering responses. The weak keypoints with $v_i=1$ can be constrained by neighboring components, whereas semantically invalid responses with $v_i=0$ are excluded from the structural representation. Overall, the proposed representation transforms SAR aircraft modeling from unordered local scattering responses into a structured physical-semantic graph, providing the foundation for the subsequent research.

\section{Method}
\label{sec:method}
\subsection{Task Definition and Overview}
\label{section3-1}

\begin{figure*}[!t]
\centering
\includegraphics[width=0.9\textwidth]{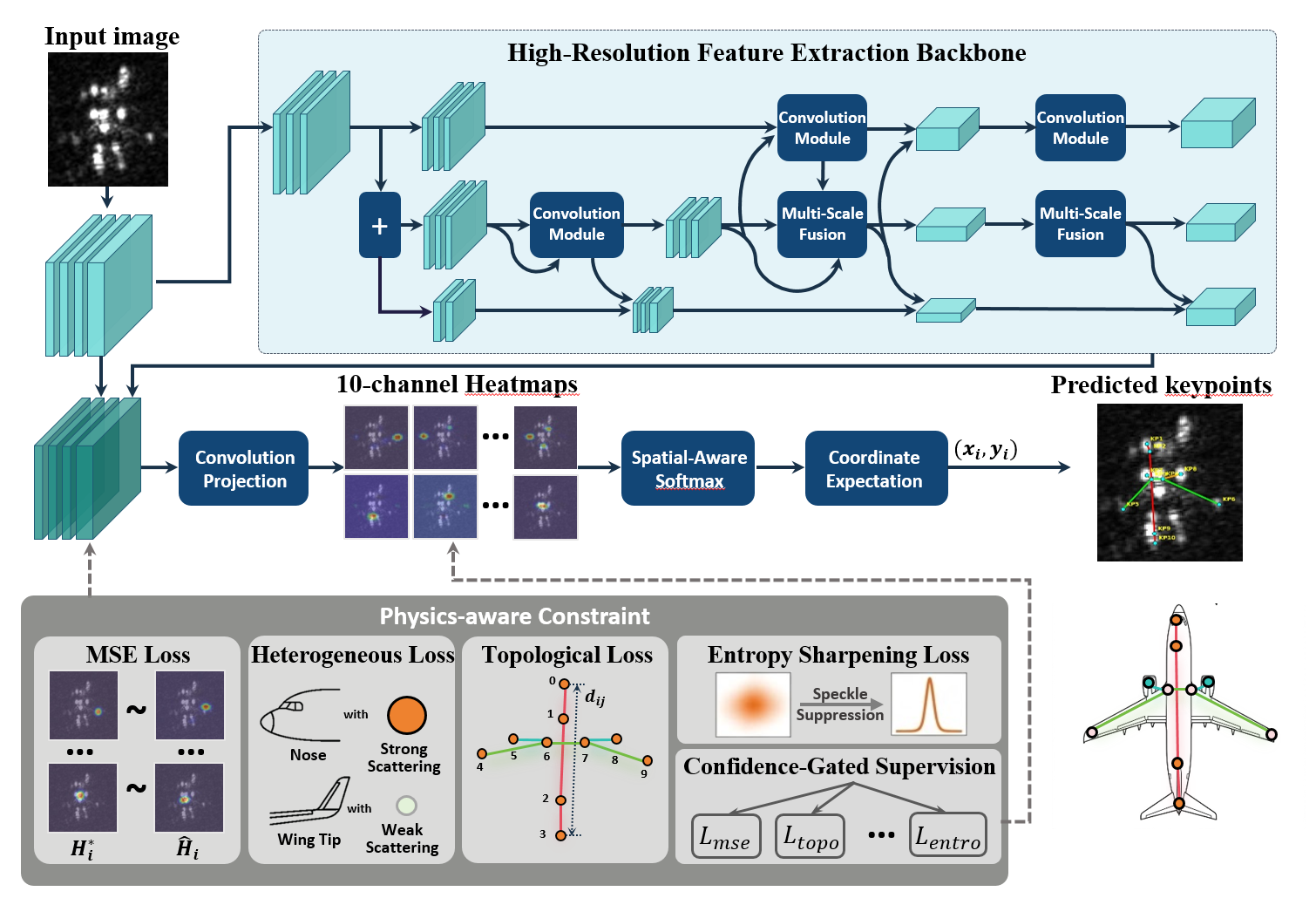}
\caption{Overall framework of the proposed S$^3$U-SAR, including high-resolution feature extraction, 10-channel heatmap prediction, spatial-aware Softmax, coordinate expectation, semantic scattering keypoint localization, and physics-aware joint supervision.}
\label{fig:framework}
\end{figure*}

Given a SAR aircraft image, the objective of S$^3$U-SAR is to localize semantic scattering keypoints and infer a structurally consistent representation guided by the physics-constrained topology defined in Section~\ref{related work}. The overall framework is illustrated in Fig.~\ref{fig:framework}. Different from conventional scattering-center extraction that focuses on local bright responses, the proposed task requires each predicted point to correspond to a physically meaningful aircraft component and to remain consistent with the global rigid-body structure.

For an input SAR image $I \in \mathbb{R}^{H \times W}$, the target representation is defined as
\begin{equation}
    \mathcal{K} = \{(k_i, v_i)\}_{i=1}^{K}, \quad K=10,
\end{equation}

where $k_i=(x_i,y_i)$ denotes the 2-D image-plane coordinates of the $i$-th semantic scattering keypoint, and $v_i$ denotes its visibility-aware scattering attribute. The goal is to predict the coordinate set $\hat{\mathcal{K}}=\{\hat{k}_i\}_{i=1}^{K}$ while preserving the semantic and structural consistency of the aircraft target.

SAR aircraft components are often represented by isolated and spatially compact scattering responses. Excessive downsampling may therefore cause irreversible loss of localization precision, especially for weakly observable components. To preserve high-resolution spatial details, we adopt HRNet \cite{Sun2019HRNet} as the backbone:

\begin{equation}
    F_{\text{feat}} = F_{\text{backbone}}(I),
\end{equation}

where $F_{\text{feat}}$ denotes the multi-scale fused feature representation extracted by the backbone. The network maintains high-resolution representations throughout feature extraction and exchanges information with parallel low-resolution branches, which is beneficial for localizing sparse scattering responses under complex SAR backgrounds.

A convolutional prediction head $\psi(\cdot)$ is then used to generate a $K$-channel heatmap:
\begin{equation}
   \hat{H} = \psi(F_{\text{feat}})=\{\hat{H}_i\}_{i=1}^{K}, 
   \quad \hat{H}_i \in \mathbb{R}^{H \times W}.
\end{equation}

Each channel $\hat{H}_i$ encodes the spatial response of a specific semantic scattering keypoint. The ground-truth heatmap $H_i^*$ is generated by applying a 2-D Gaussian kernel centered at the annotated keypoint coordinate $k_i^*$ \cite{Xiao2018SimpleBaseline}.

The standard heatmap supervision is formulated as
\begin{equation}
    \mathcal{L}_{\text{mse}} =
    \frac{1}{K}\sum_{i=1}^{K}
    \|\hat{H}_i-H_i^*\|_F^2,
\end{equation}

where $\|\cdot\|_F$ denotes the Frobenius norm. Although this pixel-domain loss provides stable coarse localization supervision, it treats all keypoints with uniform importance and does not explicitly model SAR-specific scattering heterogeneity or rigid-body topology. Therefore, additional physics-guided constraints are introduced in the following subsections.

To enable subsequent coordinate-domain physical constraints, the predicted heatmap is further decoded into continuous keypoint coordinates through a differentiable coordinate decoding operator. Specifically, each heatmap $\hat{H}_i$ is first normalized into a spatial probability distribution $\mathbf{W}_i=\{w_{t,i}\}_{t\in\Omega}$, where $\Omega$ denotes the image lattice. The coordinate of the $i$-th semantic scattering keypoint is then obtained as the expectation over all pixel positions:
\begin{equation}
    \hat{k}_i = \mathcal{D}(\hat{H}_i)
    = \sum_{t\in\Omega} w_{t,i} p_t ,
\end{equation}

where $p_t$ denotes the coordinate vector of pixel $t$, and $\mathcal{D}(\cdot)$ denotes the coordinate decoding operator. This differentiable decoding process bridges pixel-domain heatmap supervision and coordinate-domain physical regularization, allowing scattering-aware localization and topology-aware constraints to be imposed on the predicted semantic keypoints.

\subsection{Scattering Intensity Heterogeneity-Aware Localization}

The scattering intensity of different aircraft components varies significantly in SAR images. Specifically, components with dihedral or strong reflective structures, such as the nose tip and wing tips, often appear strong scattering responses, whereas other components such as fuselage may show much weaker. However, $\mathcal{L}_{\text{mse}}$ implicitly assumes uniform feature saliency across all semantic keypoints. As a result, the optimization process may be dominated by stronger scatterers, while weaker but physically meaningful components receive insufficient localization guidance.

\begin{figure}[!t]
\centering
\includegraphics[width=0.98\columnwidth]{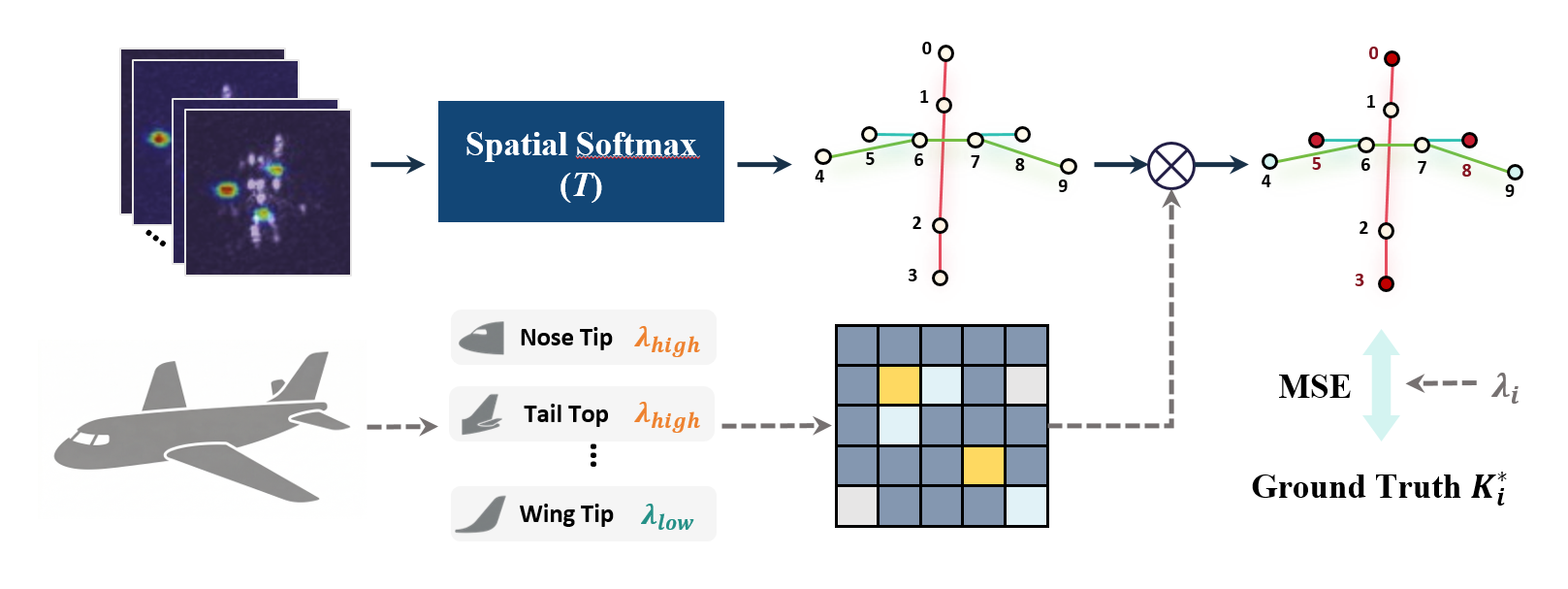}
\caption{Illustration of the scattering-intensity heterogeneity-aware localization with spatial-aware Softmax and adaptive kernel parameters $\lambda_i$ for different semantic keypoints.}
\label{fig:heterogeneous_loss}
\end{figure}

The overall process of scattering-intensity heterogeneity-aware localization is illustrated in Fig.~\ref{fig:heterogeneous_loss}. During the early stages of training, heatmap responses are often random. The extraction of peak coordinates from the heatmap is non-differentiable, causing that certain abnormally high-value points may dominate the probability distribution. To break through the limitations of pixel-domain optimization, a Spatial-Aware Softmax operation with temperature coefficient $T$ is introduced. The heatmap are transformed into normalized two-dimensional spatial probability distributions \cite{Sun2018IntegralPose}. 

Specifically, the $i$-th keypoint, let $s_{t,i}=\hat{H}_i(t)$ denote the predicted response at pixel $t$. A temperature-modulated spatial Softmax is applied:
\begin{equation}
    w_{t,i} =
    \frac{\exp(s_{t,i}/T)}
    {\sum_{j\in\Omega}\exp(s_{j,i}/T)},
\end{equation}

where $\Omega$ denotes the image lattice and $T$ is the temperature coefficient. The predicted coordinate is then obtained as the expectation over all pixel positions:
\begin{equation}
    \hat{k}_i =
    \sum_{t\in\Omega} w_{t,i}\,p_t,
\end{equation}

where $p_t$ is the coordinate vector of pixel $t$. The soft-argmax operation converts discrete heatmap localization into differentiable continuous coordinate regression. During the early training stage, a relatively large temperature is used to obtain smooth probability distributions and reduce the influence of spurious local peaks.The temperature is gradually decreased during training, so that the probability mass is progressively concentrated around reliable response regions.

To further address the optimization imbalance caused by scattering-intensity heterogeneity, we introduce a scattering-intensity-aware localization objective. Instead of applying a standard $L_1$ or $L_2$ penalty uniformly to all keypoints, we assign an adaptive kernel parameter $\lambda_i$ according to the visibility-aware scattering attribute. Highly observable keypoints with $v_i=2$ are assigned a larger $\lambda_{\text{high}}$, whereas weakly observable keypoints with $v_i=1$ are assigned a smaller $\lambda_{\text{low}}$. The heterogeneous localization loss is defined as:

\begin{equation}
    \mathcal{L}_{\text{hetero}} =
    \frac{1}{K}\sum_{i=1}^{K}
    \left(
    1-\exp\left(-\lambda_i\|\hat{k}_i-k_i^*\|_1\right)
    \right).
\end{equation}

This exponential formulation imposes a sharper penalty on scattering-salient keypoint, encouraging pixel-level alignment with salient scattering anchors. For scattering-degraded keypoint, the flatter kernel reduces over-penalization caused by unreliable local evidence and allows the model to exploit broader structural context during optimization.

\subsection{Rigid-Body Manifold Topological Constraint}

Coordinate regression based solely on independent local responses is unreliable in SAR images, where speckle noise and severe background interference easily mask true component responses. Since aircraft are rigid-body objects, their semantic scattering keypoints should satisfy stable geometric relationships determined by the physical aircraft structure. Therefore, a topology-aware constraint is introduced to guide the predicted keypoints toward a physically plausible structure defined by Eq. \ref{equation4}.

However, directly applying topological constraints to all predicted responses is inappropriate for SAR images. As discussed in Section~\ref{sec: visibility}, the visibility attribute $v_i$ decouples physical existence from scattering observability. Scattering-salient keypoints $(v_i=2)$ provide reliable local anchors, while scattering-degraded keypoints $(v_i=1)$ physically belong to the aircraft structure despite weak or unreliable scattering evidence. In contrast, semantically invalid responses $(v_i=0)$ may correspond to speckle noise, layover, multi-path artifacts, or background clutter, and should not participate in structural optimization.

Accordingly, we construct a visibility-guided valid edge set from the predefined topology:
\begin{equation}
    \mathcal{M}
    =
    \left\{
    e_{ij}=(i,j)
    \mid
    e_{ij}\in\mathcal{E},\;
    v_i\geq 1,\;
    v_j\geq 1
    \right\},
\end{equation}

where $e_{ij}$ denotes a structural edge connecting the $i$-th and $j$-th semantic scattering keypoints. $v_i$ and $v_j$ represents their corresponding visibility attributes.

An edge is retained only when both endpoints represent physically existing aircraft components. Therefore, structural relations involving either scattering-salient keypoints or scattering-degraded keypoints are preserved in the rigid-body topology. Meanwhile, edges connected to semantically invalid responses are removed, preventing clutter-induced bright artifacts from distorting the global structure. In this way, the visibility attribute defined in the representation stage is incorporated into topology regularization as a physical-semantic validity criterion.

For each valid edge $(i,j)\in\mathcal{M}$, the predicted and reference structural distances are computed as
\begin{equation}
    \hat{d}_{ij}=\|\hat{k}_i-\hat{k}_j\|_2,
    \quad
    d_{ij}^{*}=\|k_i^{*}-k_j^{*}\|_2.
\end{equation}

The rigid-body manifold topological constraint is then formulated as
\begin{equation}
    \mathcal{L}_{\text{topo}} =
    \frac{1}{|\mathcal{M}|}
    \sum_{(i,j)\in\mathcal{M}}
    \left(
    1-\exp\left(-\gamma(\hat{d}_{ij}-d_{ij}^{*})^2\right)
    \right),
\end{equation}

where $\gamma$ controls the sensitivity of the structural penalty. The exponential kernel tolerates moderate projection-induced geometric variations while assigning stronger penalties to structural outliers that violate the aircraft rigid-body manifold. Consequently, the proposed constraint allows scattering-degraded keypoints to be guided by neighboring physically connected components, while preventing semantically invalid bright responses from corrupting the structural representation.

\subsection{Speckle Suppression via Entropy Sharpening}

When the predicted heatmap is converted into the spatial probability distribution $\mathbf{W}_i$ by the spatial-aware Softmax, speckle noise and background clutter may induce multiple competing response peaks in non-target regions. Consequently, multiple competing peaks may appear in the spatial probability distribution, making the expected coordinate $\hat{\mathbf{k}}_i$ unreliable.

To reduce the uncertainty, we formulate an entropy-based regularization term directly on the normalized spatial probability matrix $\mathbf{W}_i=\{w_{t,i}\}$ at the network output. Specifically, Shannon entropy is used to measure the uncertainty of the spatial distribution \cite{Grandvalet2005EntropyMinimization}:
\begin{equation}
    \mathcal{L}_{\text{entropy}} = 
    - \frac{1}{K} \sum_{i=1}^{K} \sum_{t \in \Omega} 
    w_{t,i} \log(w_{t,i} + \epsilon),
\end{equation}

where $\epsilon$ is a small constant for numerical stability. Through entropy minimization, the spatial probability distribution is encouraged to exhibit lower uncertainty, thereby reducing the influence of speckle interference on coordinate expectation. In this way, the network is guided to concentrate probability mass around physically meaningful semantic scattering keypoints.

\subsection{Confidence-Gated Joint Supervision Strategy}

\begin{figure}[!t]
\centering
\includegraphics[width=\columnwidth]{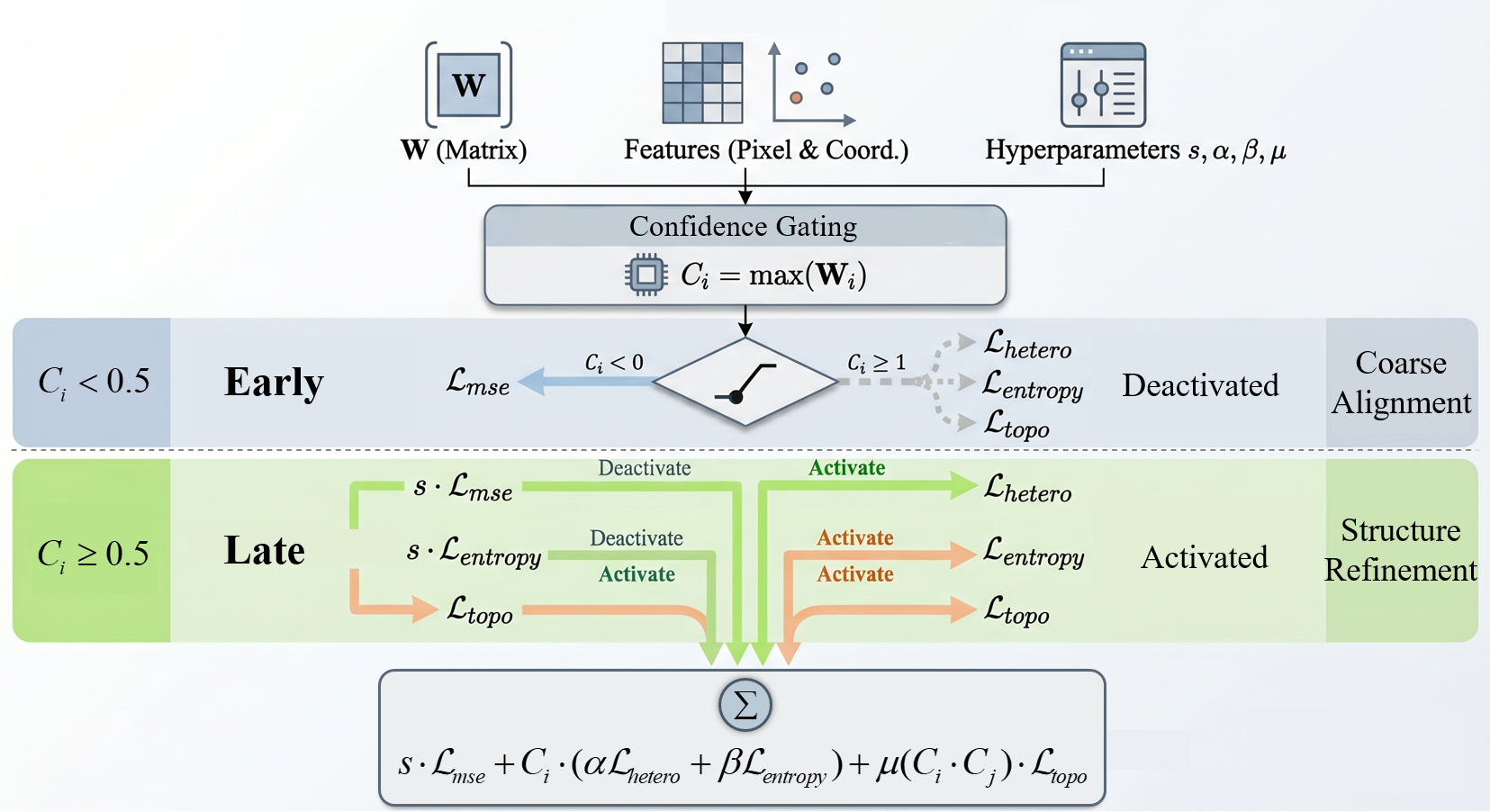}
\caption{Illustration of the confidence-gated joint supervision strategy for adaptive activation of heatmap supervision and coordinate-domain physical constraints.}
\label{fig:confidence_gating}
\end{figure}

Directly combining the pixel-domain baseline loss $\mathcal{L}_{mse}$ with the aforementioned coordinate-domain physical constraints may lead to unstable optimization due to scale mismatch and prediction uncertainty \cite{Kendall2018MultiTaskUncertainty, Yu2020PCGrad}. In the early training stage, heatmap responses are usually dispersed, and the decoded coordinates are not sufficiently reliable. Under this condition, enforcing scattering-aware localization and topological regularization may introduce noisy gradients and interfere with coarse spatial learning.

To address the issue, a Confidence-Gated Joint Supervision strategy is proposed, as illustrated in Fig.~\ref{fig:confidence_gating}. For the $i$-th semantic scattering keypoint, the maximum response in its spatial probability distribution is used as a dynamic localization confidence score $C_i = \max(\mathbf{W}_i)$. A larger $C_i$ indicates a sharper and more reliable spatial response, whereas a smaller $C_i$ reflects uncertain localization. With a scale alignment coefficient $s$, the total training objective is formulated as:

\begin{equation}
\begin{aligned}
    \mathcal{L}_{\text{total}} =
    &\, s \cdot \mathcal{L}_{\text{mse}}
    + \sum_{i=1}^{K} C_i
    \left(
    \alpha \mathcal{L}_{\text{hetero}}^{i}
    +
    \beta \mathcal{L}_{\text{entropy}}^{i}
    \right)  \\
    &+ \mu
    \sum_{e_{ij}\in \mathcal{M}}
    C_i C_j \cdot \mathcal{L}_{\text{topo}}^{ij}.
\end{aligned}
\end{equation}

where $\mathcal{L}_{\text{hetero}}^{i}$ and $\mathcal{L}_{\text{entropy}}^{i}$ denote the keypoint-wise scattering-intensity-aware localization loss and entropy regularization term, respectively. $\mathcal{L}_{\text{topo}}^{ij}$ denotes the edge-wise topological constraint for the valid structural edge $e_{ij}$, and $\alpha$, $\beta$, and $\mu$ balance the contributions of different physical constraints.

The strategy enables an adaptive optimization process. In the early stage, when the predicted distributions are uncertain and $C_i$ remains small, the coordinate-domain constraints are automatically suppressed, allowing the network to focus on stable coarse localization through $\mathcal{L}_{\text{mse}}$. As training proceeds and reliable response peaks gradually emerge, the physical constraints are progressively activated. Consequently, scattering-aware refinement and topology-consistent structural optimization can be introduced in a confidence-dependent manner, reducing gradient conflicts between early heatmap learning and late-stage physical regularization.

\section{Experiment}
\label{experiment}
In this section, comprehensive experiments are conducted to validate the effectiveness of our method. First, a fine-grained, precisely annotated dataset for SAR aircraft targets is constructed. Both qualitative and quantitative evaluations are conducted to compare the proposed approach against state-of-the-art keypoint extraction methods. Subsequently, comprehensive ablation studies are performed to verify the effectiveness of the core components. To further assess the robustness and generalization capability of the model, open-set cross-category experiments are designed, wherein the network is trained on known aircraft categories and evaluated on unseen categories. In addition, cross-dataset analyses are carried out to demonstrate the transferability of the method on untrained data. Finally, the potential of the proposed method for downstream applications is validated by the results of aircraft orientation estimation.

\subsection{Dataset Construction and Specifications}
To evaluate the effectiveness of the proposed S$^3$U-SAR method and advance research in fine-grained representation of SAR aircraft targets, a publicly available dataset with fine-grained annotations, named KP-SAR-Aircraft-1.0, is constructed. The images are acquired by the Gaofen-3 satellite with a spatial resolution of 1 m \cite{Sun2022SCAN, Wang2023SARAIRcraft}. The dataset comprises 2,990 samples of seven specific aircraft categories, including A220, A320/321, A330, ARJ21, Boeing 737, Boeing 787, and others. Based on the actual physical structures as illustrated in Section \ref{related work}, the key components of the targets are precisely annotated, and each keypoint is assigned a visibility semantic attribute. Furthermore, topological structural relationships are established according to the component category of each scattering center. To facilitate downstream applications and future studies, the category and heading orientation information are additionally annotated. 

\subsection{Implementation Details}
In the experiments, HRNet-w32 \cite{Sun2019HRNet} is adopted as the backbone, which is pretrained on the ImageNet dataset. The input images are resized to 256 x 192. The model is trained for 20 epochs using the Adam optimizer \cite{Kingma2015Adam} with an initial learning rate of 1e-3. Regarding the data augmentation strategy, only rotation and contrast enhancements are employed to accommodate the fixed rigid geometric structure of aircraft targets. Meanwhile, conventional operations such as scaling, random cropping, and flipping are strictly prohibited to avoid disrupting the spatial topological relationships and electromagnetic scattering characteristics of the targets. Following the introduction of the confidence-gated strategy, the temperature coefficient $T$ is initially set to 1.0 and gradually decayed to 0.3. The hyperparameters are empirically set as $\alpha=0.8$, $T_{\mu}=0.4$, and $\beta=0.15$. The entire framework is implemented based on the MMPose \cite{MMPose2020}, and all experiments are conducted on 4 NVIDIA RTX A6000 GPUs.

\begin{table*}[!t]
\centering
\caption{Keypoint estimation results compared with state-of-the-art methods}
\label{tab:sota_comparison}
\scalebox{1.2}{
\begin{tabular}{llcccccccc}
\toprule[1.5pt]
Method & Backbone & $\text{AP}$ & $\text{AP}_{50}$ & $\text{AP}_{75}$ 
& $\text{AP}_{\text{M}}$ & $\text{AP}_{\text{L}}$ 
& $\text{AR}$ & $\text{AR}_{50}$ & $\text{AR}_{75}$ \\
\midrule[1pt]

\multirow{2}{*}{HRNet\cite{Sun2019HRNet}}
& HRNet-W32  & 54.7 & 81.9 & 61.6 & 54.3 & 68.4 & 68.4 & 89.8 & 75.6 \\
& HRNet-W48  & 55.3 & 82.8 & 61.8 & 54.6 & 68.4 & 69.1 & 90.1 & 77.0 \\

\midrule
\multirow{2}{*}{SimpleBaseline\cite{Xiao2018SimpleBaseline}}
& ResNet-50  & 50.8 & 79.9 & 54.0 & 49.9 & 67.2 & 66.9 & 89.5 & 73.1 \\
& ResNet-101 & 52.3 & 81.8 & 55.8 & 51.7 & 65.1 & 66.9 & 89.8 & 72.9 \\

\midrule
\multirow{3}{*}{Spatial-Aware Reg.\cite{Wang2024SpatialAware}}
& HRNet-W48  & 51.3 & \underline{84.2} & 51.8 & 50.6 & 57.3 & 60.5 & 89.7 & 64.3 \\
& ResNet-34  & 42.8 & 75.2 & 45.7 & 41.2 & 58.3 & 59.6 & 86.6 & 67.1 \\
& ResNet-50  & 47.9 & 83.6 & 48.7 & 47.1 & 55.3 & 57.9 & 88.3 & 62.0 \\

\midrule
\multirow{2}{*}{ViTPose \cite{Xu2022ViTPose}}
& Base  & 55.5 & 83.3 & 61.7 & 54.7 & 67.3 & 69.5 & 90.2 & 76.6 \\
& Small & 55.4 & 82.2 & 61.5 & 55.4 & 67.7 & 69.2 & 89.9 & 76.2 \\

\midrule
\multirow{2}{*}{SimCC \cite{Li2022SimCC}}
& SimDR  & 36.7 & 75.0 & 32.4 & 35.9 & 51.7 & 52.1 & 85.0 & 54.7 \\
& SimDR* & 53.4 & 83.6 & 58.9 & 52.6 & 66.2 & 67.0 & 90.3 & 74.4 \\

\midrule
DiffusionPose \cite{Wang2026DiffusionPose} & -- & 36.2 & 74.0 & 32.8 & 36.2 & 48.4 & 51.0 & 84.4 & 53.8 \\

\midrule
ProbPose \cite{Purkrabek2025ProbPose} & -- & 46.7 & 82.9 & 31.0 & 39.1 & 41.9 & 48.3 & 86.2 & 47.1 \\

\midrule[1pt]
\multirow{5}{*}{S$^3$U-SAR}
& ViTPose    & 55.5 & 81.7 & \underline{63.9} & 54.8 & 65.8 & 69.4 & 89.6 & \underline{78.4} \\
& HRNet-W48  & \underline{56.7} & 83.9 & 63.8 & \underline{56.0} & \underline{69.5} & \underline{69.7} & 90.1 & 77.1 \\
& ResNet-50  & 52.7 & 83.3 & 57.1 & 52.4 & 65.1 & 67.1 & 90.2 & 73.9 \\
& ResNet-101 & 54.3 & 83.7 & 58.9 & 53.5 & 66.6 & 68.0 & \underline{90.9} & 74.7 \\
& HRNet-W32  & \textbf{59.3} & \textbf{85.4} & \textbf{66.8} & \textbf{58.4} & \textbf{73.0} & \textbf{71.9} & \textbf{91.6} & \textbf{79.8} \\

\bottomrule[1.5pt]
\end{tabular}
}
\end{table*}

\subsection{Comparisons with Baseline Methods}
To evaluate the effectiveness of the proposed S$^3$U-SAR, extensive comparative experiments are conducted on the newly constructed KP-SAR-Aircraft-1.0 dataset. To the best of our knowledge, S$^3$U-SAR is the first work dedicated to extracting key semantic scattering centers of SAR aircraft targets. In the absence of direct in-domain counterparts, we select representative approaches from the 2D human pose estimation domain as our baselines, including HRNet\cite{Sun2019HRNet}, Spatial-Aware Regression\cite{Wang2024SpatialAware}, ViTPose\cite{Xu2022ViTPose}, SimCC\cite{Li2022SimCC}, ProbPose\cite{Purkrabek2025ProbPose}, and DiffusionPose\cite{Wang2026DiffusionPose}. For a comprehensive evaluation, we adopt the standard metric suite \cite{Lin2014COCO}, including $\text{AP}$, $\text{AP}_{50}$, $\text{AP}_{75}$, $\text{AP}_{M}$, $\text{AP}_{L}$, $\text{AR}$, $\text{AR}_{50}$, and $\text{AR}_{75}$. The quantitative comparison results are summarized in Table~\ref{tab:sota_comparison}.

As observed from the results, S$^3$U-SAR achieves an overall average precision ($\text{AP}$) of 59.3\%, outperforming the mainstream HRNet-W48 and the transformer-based ViTPose-base by 4.0\% and 3.8\%, respectively. Notably, the improvement of over 5.0\% on $\text{AP}_{75}$ metric particularly highlights its superior high-fidelity localization capability. The performance further confirm that the incorporation of scattering characteristics and rigid topological constraints significantly improves high-precision localization. Furthermore, the consistently superior performance across different backbone architectures further validates the effectiveness and robustness of our approach.

\begin{figure*}[!t]
\centering
\includegraphics[width=\textwidth]{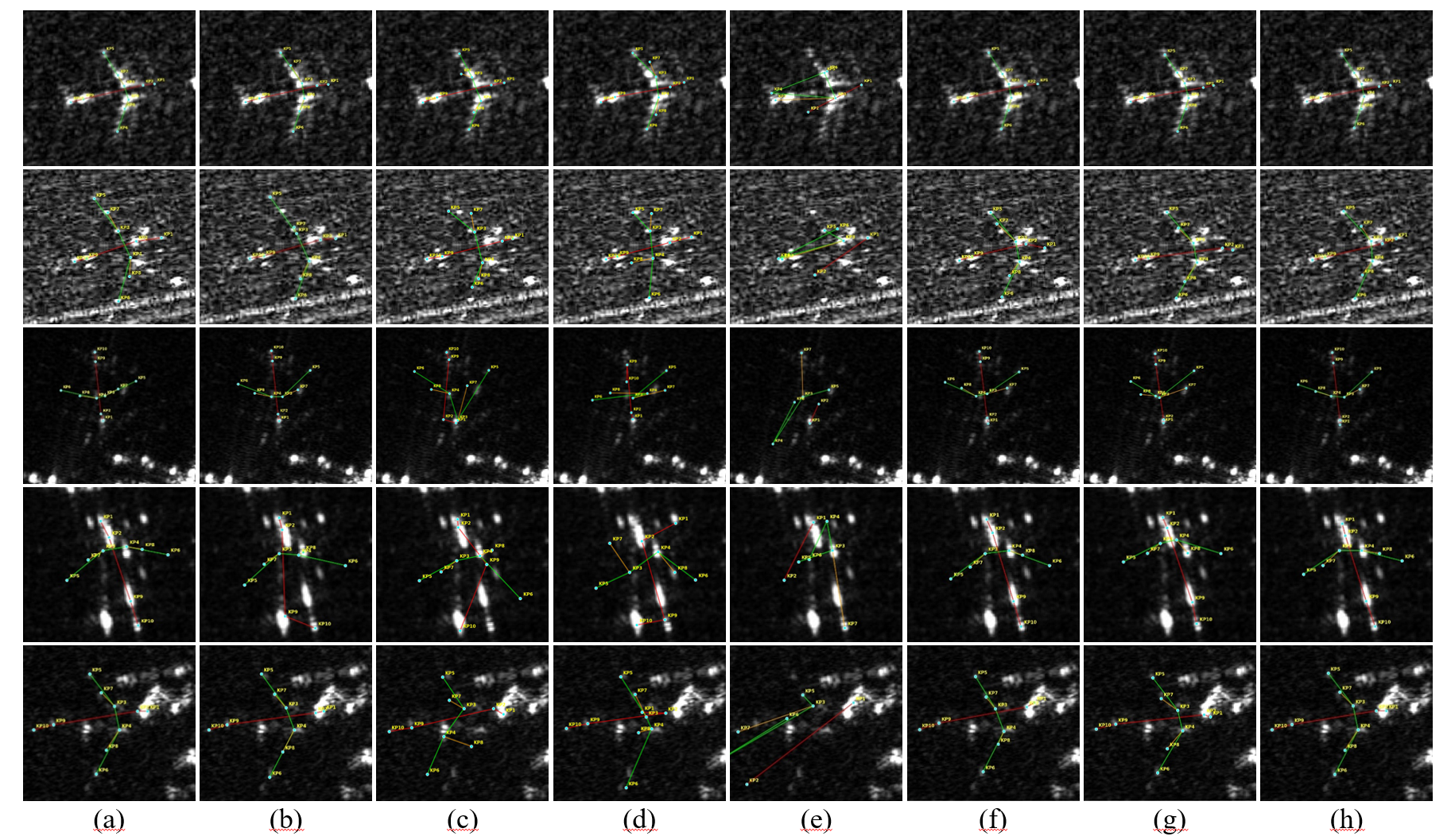}
\caption{Qualitative comparison of semantic scattering keypoint localization results of different methods. (a) GT. (b) ViTPose. (c) SimCC. (d) DiffusionPose. (e) ProbPose. (f) HRNet. (g) Spatial-Aware Regression. (h) S$^3$U-SAR.}
\label{fig:qualitative_comparison}
\end{figure*}

Fig.~\ref{fig:qualitative_comparison} further presents qualitative visual comparisons of keypoint localization to complement the quantitative results. Visual analysis reveals that generic keypoint localization models, such as ProbPose and SimCC, are highly sensitive to the coherent speckle noise and self-occlusion specific to SAR data. These methods suffer from severe topological distortion and local feature degradation. Furthermore, purely data-driven models like DiffusionPose, ViTPose, and HRNet tend to extract generic optical features, causing their predictions to deviate from the actual aircraft structures, failing to align with the true high-intensity scattering regions. Similarly, Spatial-Aware Regression, which relies on direct point regression, is susceptible to speckle noise. In contrast, constrained by the physical-electromagnetic semantic joint representation, our method achieves robust and precise localization even under complex background interference.

\begin{table}[!t]
\centering
\caption{Ablation study of different constraint terms.}
\label{tab:ablation_constraints}
\renewcommand{\arraystretch}{1.25}
\resizebox{\columnwidth}{!}{
\begin{tabular}{cccccccccc}
\toprule[1.5pt]
$\mathcal{L}_{\text{mse}}$ & 
$\mathcal{L}_{\text{hetero}}$ & 
$\mathcal{L}_{\text{topo}}$ & 
$\mathcal{L}_{\text{entropy}}$ & 
$\text{AP}$ & $\text{AP}_{50}$ & $\text{AP}_{75}$ & 
$\text{AR}$ & $\text{AR}_{50}$ & $\text{AR}_{75}$ \\
\midrule[1pt]
$\checkmark$ &  &  &  & 55.7 & 82.0 & 61.4 & 69.6 & 89.8 & 76.7 \\
$\checkmark$ & $\checkmark$ &  &  & 56.7 & 83.6 & 62.9 & 70.2 & 90.8 & 77.2 \\
$\checkmark$ &  & $\checkmark$ &  & \underline{58.2} & \underline{85.3} & 65.0 & \underline{70.9} & \textbf{91.6} & 78.0 \\
$\checkmark$ &  &  & $\checkmark$ & 56.4 & 83.3 & 63.0 & 69.8 & 90.7 & 77.7 \\
$\checkmark$ & $\checkmark$ & $\checkmark$ &  & 57.7 & 85.1 & \underline{65.4} & 70.8 & 91.4 & \underline{78.3} \\
$\checkmark$ & $\checkmark$ &  & $\checkmark$ & 57.0 & 83.5 & 64.1 & 70.2 & 90.8 & 77.3 \\
$\checkmark$ &  & $\checkmark$ & $\checkmark$ & 57.9 & 85.1 & 65.2 & 70.8 & \textbf{91.6} & 78.2 \\
$\checkmark$ & $\checkmark$ & $\checkmark$ & $\checkmark$ & 
\textbf{59.3} & \textbf{85.4} & \textbf{66.8} & 
\textbf{71.9} & \textbf{91.6} & \textbf{79.8} \\
\bottomrule[1.5pt]
\end{tabular}
}
\end{table}

\subsection{Ablation Study}
To verify the individual contributions of the proposed loss constraints, detailed ablation results are presented in Table \ref{tab:ablation_constraints}. The model optimized solely with the mean squared error loss $\mathcal{L}_{mse}$ is established as the baseline, yielding $\text{AP}$ of $55.7\%$ and $\text{AP}_{75}$ of $61.4\%$. Separate integration of the physical and geometric constraints into the baseline leads to consistent performance improvements. Notably, the performance gain contributed by the topological structural constraint $\mathcal{L}_{topo}$ alone is the most prominent, achieving the improvement of $2.5\%$ and $3.6\%$ in $\text{AP}$ and $\text{AP}_{75}$, respectively. Furthermore, ablation studies show that the proposed constraint terms are complementary and jointly improve localization performance. Consequently, optimal performance is attained when all constraints are employed. The findings demonstrate that the incorporation of multi-dimensional physical constraints effectively drives the network to converge toward a superior parameter space.

\begin{figure}[!t]
\centering
\includegraphics[width=\columnwidth]{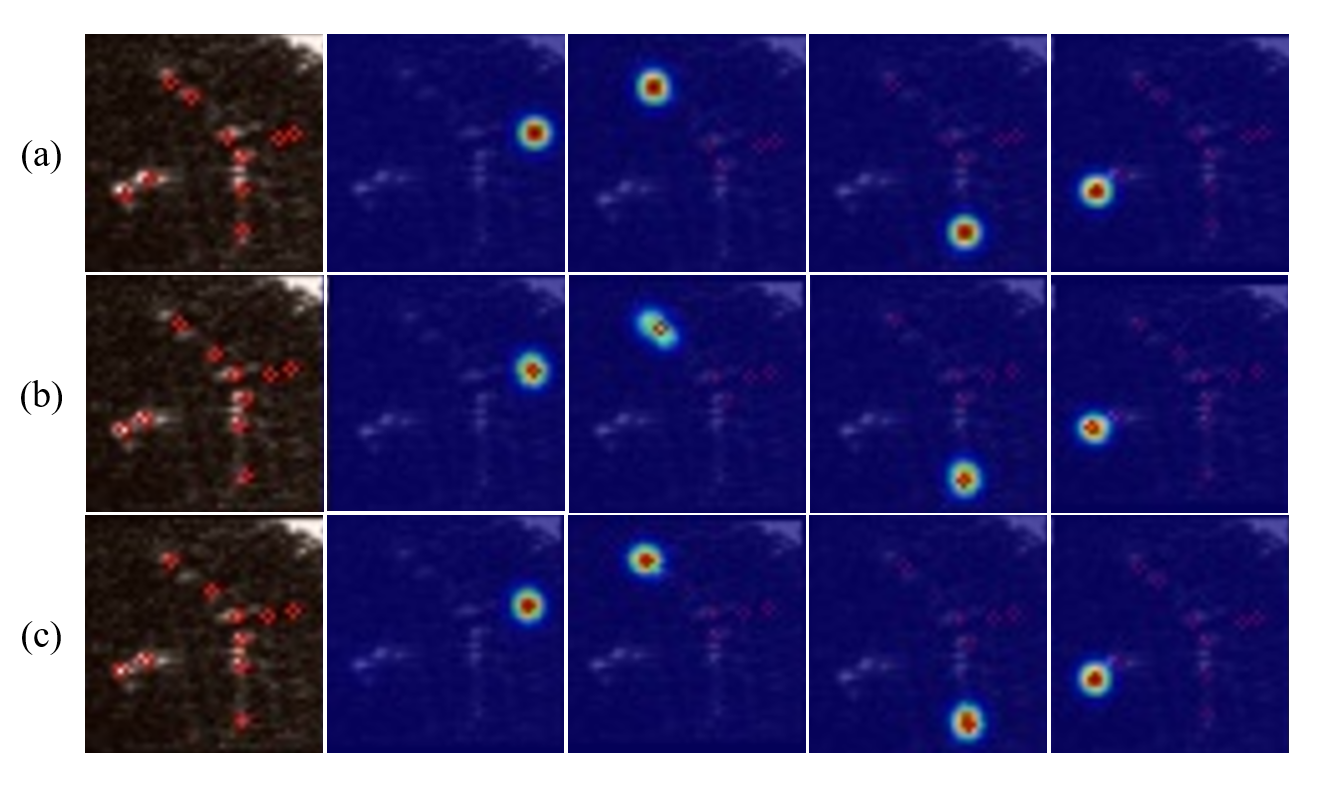}
\caption{Visualization of heatmap responses for scattering-intensity heterogeneity-aware localization. (a) GT. (b) w/o $\mathcal{L}_{\text{hetero}}$. (c) w/ $\mathcal{L}_{\text{hetero}}$.}
\label{fig:ablation_hetero}
\end{figure}

\subsubsection{Influence of Scattering Intensity Heterogeneity-Aware Localization}
As shown in Table \ref{tab:ablation_constraints}, the introduction of $\mathcal{L}_{here}$ yields an absolute improvement of $1.0\%$ in $\text{AP}$. To thoroughly investigate the underlying mechanism, a qualitative analysis of the feature response heatmaps is conducted, with results visualized in Fig.~\ref{fig:ablation_hetero}. In the predictions of the baseline model, the scattering response of aircraft components in the SAR images exhibit highly discrete and non-uniform characteristics. Consequently, the response regions suffer from energy dispersion, which causes the predicted peaks to severely deviate from the physical centers. Conversely, the injection of the scattering heterogeneity-aware constraint, feature divergence within the strong scattering neighborhoods is successfully suppressed. The output heatmaps exhibit a convergent Gaussian distribution, achieving high-precision alignment with the ground truth. The results indicate that the incorporation of $\mathcal{L}_{here}$ could alleviate the gradient instability caused by non-uniform strong scattering centers, while enhancing the robustness against speckle noise.

\begin{figure}[!t]
\centering
\includegraphics[width=\columnwidth]{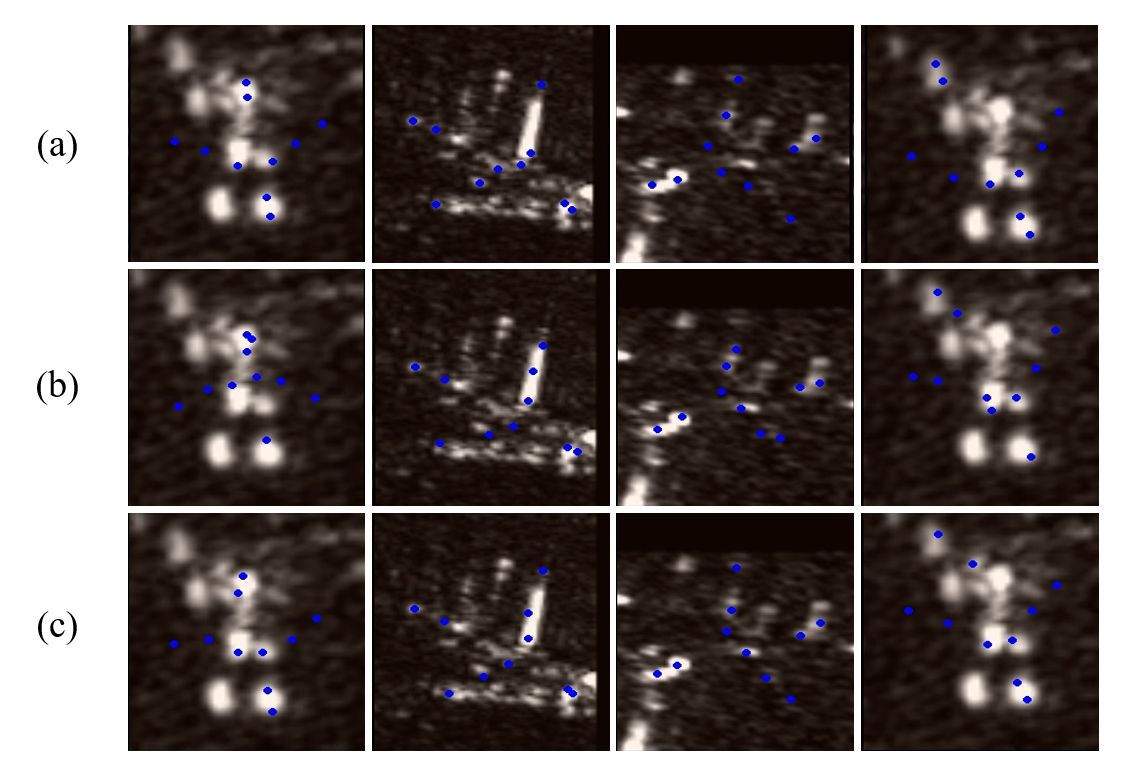}
\caption{Visualization of the topological constraint. (a) GT. (b) w/o $\mathcal{L}_{\text{topo}}$. (c) w/ $\mathcal{L}_{\text{topo}}$.}
\label{fig:ablation_topo}
\end{figure}

\subsubsection{Influence of Rigid-Body Manifold Topological Constraint}
To examine the specific effect of the rigid-body manifold topology constraint $\mathcal{L}_{topo}$, further visual analyses are conducted, with results shown in Fig.~\ref{fig:ablation_topo}. The introduction of $\mathcal{L}_{topo}$ yields absolute improvements of 2.5\% in $\text{AP}$ and 3.6\% in $\text{AP}_{75}$. It is observed that targets in SAR images exhibit a discrete distribution of scattering points. The baselines without topological constraints lack a global representation of the rigid structure. Hence, the keypoint mislocalization and topology distortions could be observed in regions. By contrast, the incorporation of $\mathcal{L}_{topo}$ constrains the predicted keypoint configuration toward a physically consistent structure. As illustrated in Fig.~\ref{fig:ablation_topo}(c), the constraint effectively reduces structural deviations and guides the predicted keypoints to conform to the rigid-body topology of the aircraft.

\subsection{Discussion of the Confidence-Gated Joint Supervision Strategy}
\begin{figure}[!t]
\centering
\includegraphics[width=0.9\columnwidth]{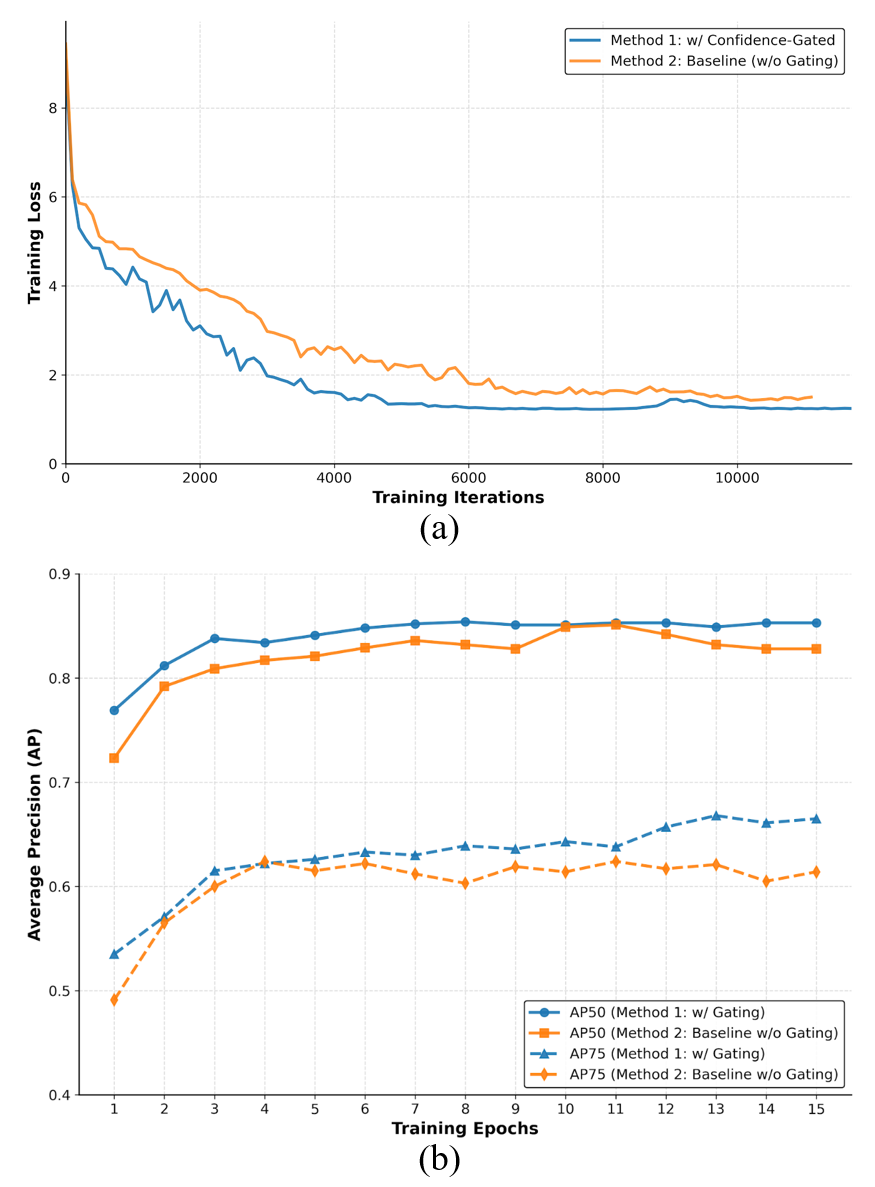}
\caption{Training dynamics of the confidence-gated joint supervision strategy. (a) The training loss comparison. (b) The precision comparison.}
\label{fig:gating_training_dynamics}
\end{figure}

To verify the effectiveness of the confidence-gated joint supervision strategy in mitigating multi-task optimization conflicts, the loss convergence and the validation accuracy during the training process are analyzed in this section. In the feature space, inherent gradient conflicts exist between coarse localization and fine structural refinement. As shown in the top panel of Fig.~\ref{fig:gating_training_dynamics}(a), the baseline suffers from delayed convergence and unstable optimization dynamics. The phenomenon indicates that the network suffers from mutual interference among different constraints. By comparison, following the introduction of the confidence-gated strategy, the smooth transition of the joint supervision process is achieved and the training loss ultimately stabilizes at a lower level.

Furthermore, to provide additional validation of the effectiveness, the curves of average precision are presented in the bottom of Fig.~\ref{fig:gating_training_dynamics} (b). A higher precision level of $\text{AP}_{50}$ is reached at an earlier stage facilitated by the gating mechanism. More crucially, the topological constraints are misguided by low-quality early predictions of the baseline. The misguidance amplifies the error loss and restricts the performance upper bound. Consequently, signs of performance fluctuation and degradation can be observed in $\text{AP}_{75}$ during the later stages of training. In contrast, the $\text{AP}_{75}$ achieve stable improvement with the introduction of the gating strategy. The results indicate that the dynamic confidence-gated strategy effectively prevents negative interference from low-confidence predictions on the joint constraints, thereby enhancing the performance upper bound.

\begin{table}[!t]
\centering
\caption{Cross-category generalization results on unseen categorys.}
\label{tab:cross_type_generalization}
\renewcommand{\arraystretch}{1.15}
\resizebox{\columnwidth}{!}{
\begin{tabular}{lcccccc}
\toprule[1.5pt]
Unseen Category & $\text{AP}$ & $\text{AP}_{50}$ & $\text{AP}_{75}$ 
& $\text{AR}$ & $\text{AR}_{50}$ & $\text{AR}_{75}$ \\
\midrule[1pt]
A330       & 51.6 & 85.3 & 55.3 & 63.4 & 90.6 & 71.9 \\
A320       & 34.8 & 71.5 & 30.9 & 51.3 & 84.4 & 55.6 \\
Boeing 737  & 38.2 & 76.5 & 32.9 & 53.9 & 87.6 & 56.2 \\
Boeing 787  & 46.6 & 81.0 & 46.5 & 61.8 & 89.7 & 68.1 \\
\bottomrule[1.5pt]
\end{tabular}
}
\end{table}

\begin{figure}[!t]
\centering
\includegraphics[width=1.0\columnwidth]{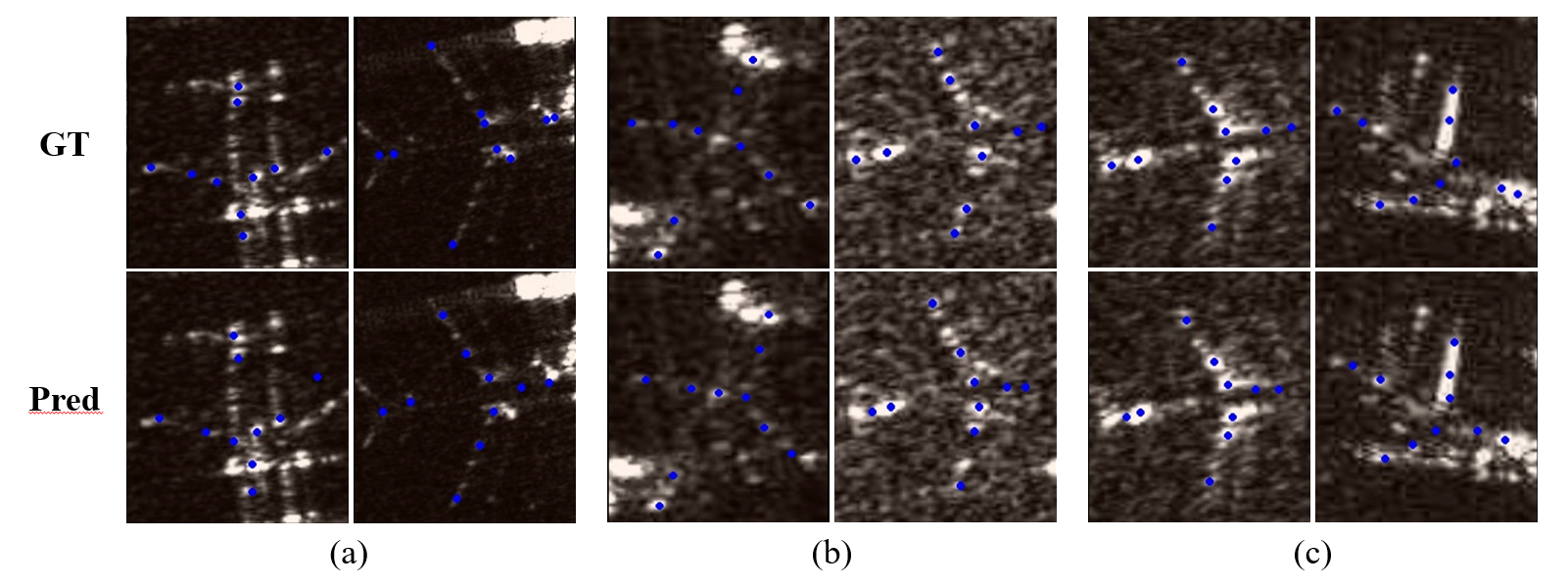}
\caption{Qualitative results under cross-category evaluation. The first row shows the ground truth, and the second row shows the predicted results. (a) A330. (b) Boeing 787. (c) Boeing 737.}
\label{fig:cross_category_visual}
\end{figure}

\subsection{Discussion on Cross-Category Generalization and Robustness}
To thoroughly investigate the generalization capability and feature extraction robustness of the proposed method when confronting unseen targets, cross-category zero-shot validation experiments are designed. Under this open-set strategy, samples from specific categories such as A330, A320, Boeing 737, and Boeing 787, are strictly excluded from the training set and are utilized exclusively for evaluation during the testing phase. As observed in Table \ref{tab:cross_type_generalization}, although the network is not exposed to any references of the target aircraft during training, highly reliable feature decoupling and localization capabilities are maintained. For large aircraft such as A330 and Boeing 787, average precisions $\text{AP}$ of 51.6\% and 46.6\% are achieved, respectively. Furthermore, the fundamental localization metric $\text{AP}_{50}$ reaches 85.3\% and 81.0\%. Even when evaluating the A320 and Boeing 737, which possess relatively fewer scattering features, an $\text{AP}_{50}$ level exceeding 71\% is robustly maintained.

The qualitative visual results of key scattering center extraction are illustrated in Fig.~\ref{fig:cross_category_visual}. On unseen samples, key scattering components are accurately localized without exhibiting any topological structural collapse. These results strongly demonstrate that the introduced joint representation mechanism enables the network to breakthrough data-driven paradigms. Rather than merely memorizing the training distribution, the model is successfully guided to learn the intrinsic correlation between electromagnetic scattering characteristics and rigid-body geometric relations. In summary, exceptional cross-domain transferability and robustness in open-set scenarios are exhibited by the proposed method for extracting key scattering centers of unseen categories.

\begin{figure}[!t]
\centering
\includegraphics[width=\columnwidth]{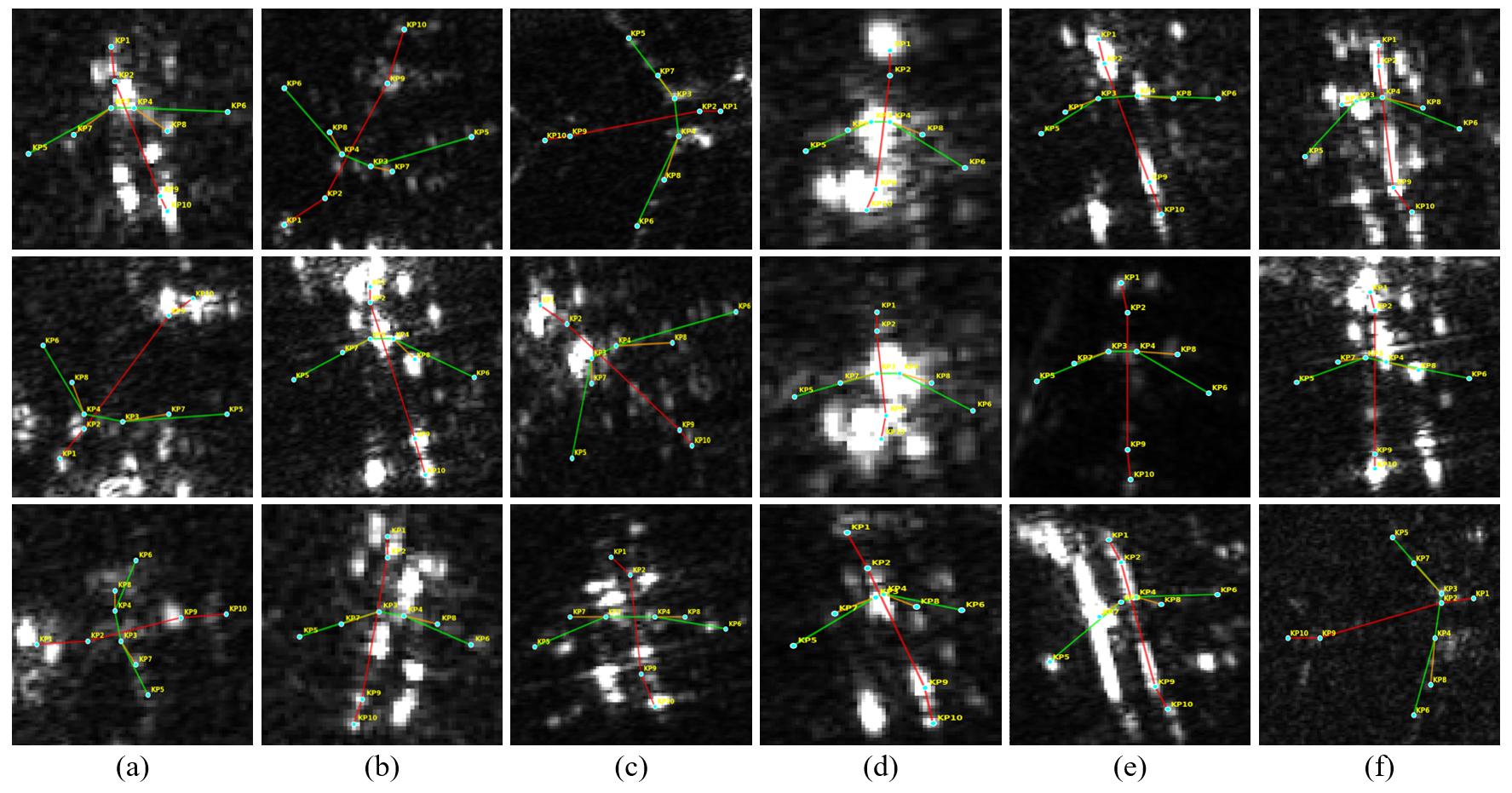}
\caption{Qualitative results under cross-dataset evaluation. (a) A220. (b) A320. (c) A330. (d) ARJ21. (e) Boeing 737. (f) Boeing 787.}
\label{fig:cross_dataset_visual}
\end{figure}

\subsection{Discussion on Cross-Dataset Domain Adaptability and Generalization}
To verify the generalization capability and robustness of the proposed method across diverse scenarios, a cross-dataset zero-shot transfer experiment is designed. Specifically, the model is trained exclusively on the KP-SAR-Aircraft-1.0 dataset and subsequently deployed for direct forward inference on the completely unseen SAR-ACD dataset \cite{Sun2022SCAN}. The corresponding qualitative visualizations are presented in Fig.~\ref{fig:cross_dataset_visual}. As observed from the results, without any fine-tuning on the target domain data, the proposed method successfully extracts the key scattering centers of various aircrafts and correctly reconstructs their topological structures. Even under the interference of coherent speckle noise and complex backgrounds, the predicted keypoints remain accurately focused on core component regions, such as the nose, tail, and engines. Furthermore, the reconstructed geometric topology strictly aligns with the underlying physical mechanisms. Notably, almost no global topological collapse or semantic drift induced by scene domain shifts is observed. These results demonstrate that the proposed electromagnetic-aware semantic representation reduces overfitting to the distribution of a single source dataset. Consequently, substantial potential for practical application in complex, open-world real scenarios is exhibited.

\begin{table}[!t]
\centering
\caption{Orientation estimation results comparison.}
\label{tab:orientation_estimation}
\renewcommand{\arraystretch}{1.15}
\resizebox{\columnwidth}{!}{
\begin{tabular}{lccccc}
\toprule[1.5pt]
Method & $\text{P}_{\text{1}^\circ}$ & $\text{P}_{\text{5}^\circ}$ & $\text{P}_{\text{15}^\circ}$ & $\text{P}_{\text{30}^\circ}$ & $\text{MAE}(^\circ)$ \\
\midrule[1pt]
ResNet-18 \cite{He2016ResNet} & 8.20  & 38.62 & 77.90 & 86.72 & 17.31 \\
ResNet-34 \cite{He2016ResNet} & 13.06 & 49.22 & \underline{81.14} & \underline{90.40} & \underline{14.79} \\
ResNet-50 \cite{He2016ResNet} & 12.95 & 46.88 & 80.58 & 88.62 & 16.31 \\
VGG-16 \cite{Simonyan2015VGG}   & \underline{14.51} & \underline{49.44} & 79.58 & 87.95 & 15.78 \\
ViT \cite{Dosovitskiy2021ViT}      & 11.72 & 40.62 & 74.67 & 86.05 & 17.37 \\
% SARATR-X \cite{Li2025SARATRX}      & X & X & X & X & X \\
Ours & \textbf{30.13} & \textbf{68.53} & \textbf{88.28} & \textbf{91.74} & \textbf{13.15} \\
\bottomrule[1.5pt]
\end{tabular}
}
\end{table}

\subsection{Performance Validation in Downstream Applications}

To further validate the practical utility of the proposed method in downstream tasks, comparative experiments are conducted on the aircraft orientation estimation task. For the baseline approaches, classic backbone networks are used for feature extraction, including ResNet~\cite{He2016ResNet}, ViT~\cite{Dosovitskiy2021ViT}, and SARATR-X~\cite{Li2025SARATRX}. A unified angle classification head is then adopted for end-to-end prediction. The proposed method follows a structure-based orientation estimation scheme. It first reconstructs the aircraft topology from the extracted semantic key scattering centers and then estimates the orientation angle according to the direction from the tail keypoint to the nose keypoint. The quantitative results are summarized in Table \ref{tab:orientation_estimation}. Mean Angle Error (MAE) indicates the Mean Absolute Error of the angle \cite{Sang2025KeypointAngle, Wang2024RLOrientation}. $\text{P}_{\text{y}^\circ}$ represents the Average Precision for the predicted angle within $\text{y}^\circ$. 

In terms of MAE, the proposed method achieves the lowest error of $13.15^\circ$, consistently outperforming all evaluated traditional vision baselines. Notably, under strict error thresholds, accuracy rates of $30.13\%$ and $68.53\%$ of $\text{P}_{1^\circ}$ and $\text{P}_{5^\circ}$ metrics are achieved, respectively, yielding absolute improvements of $17.07\%$ and $19.31\%$ percentage points over the best-performing baseline model. Instead of implicitly learning the orientation distribution from discrete strong scattering responses, the proposed method provides an explicit structure-based solution for target orientation estimation. The design reduces the dependence on distribution fitting and improves robustness to azimuth variations in SAR images. Ultimately, the results further confirm that the proposed method effectively bridges physical scattering characteristics with high-level semantic information, providing a promising approach for the future development of SAR image interpretation.

\section{Conclusion}
\label{conclusion}
In this paper, we established semantic scattering structure understanding as a new representation paradigm for SAR aircraft interpretation. Compared with conventional methods based on scattering centers, the proposed paradigm advances aircraft target representation from isolated local responses to global structure understanding. In this way, electromagnetic scattering responses are explicitly associated with physical aircraft components, weakly observable but physically existing structures are preserved, and a complete, stable semantic scattering representation is constructed.

Subsequently, we developed S³U-SAR, a physics-driven framework for semantic scattering keypoint localization and structural topology reconstruction. Physics-guided constraints are introduced to align electromagnetic characteristics with aircraft actual structure, while a confidence-gated supervision strategy is incorporated for stable optimization. We further constructed KP-SAR-Aircraft-1.0 as an initial fine-grained benchmark, providing semantic keypoints, visibility attributes, topological relations, category labels, and orientation annotations.

Comprehensive experiments, including baseline comparisons, ablation studies, cross-category and cross-dataset evaluations, and downstream orientation estimation, consistently demonstrate the effectiveness, robustness, and practical utility of the proposed approach. Future work will expand the benchmark to larger datasets, more imaging scenarios, and broader target categories for open-world SAR target understanding.

\bibliographystyle{unsrt}
\bibliography{references_1}

\end{document}